\documentclass[12pt]{iopart}
\usepackage[utf8]{inputenc}
\usepackage{textgreek}  % for references with greek symbols in them
\DeclareUnicodeCharacter{2212}{\ensuremath{-}}  % map the Unicode minus to an ASCII/math minus
\usepackage[usetitle=true]{rsc}
\newcommand{\super}[1]{$^{\textrm{#1}}$}
\newcommand{\sub}[1]{$_{\textrm{#1}}$}

\usepackage{pdftexcmds}
\expandafter\let\csname equation*\endcsname\relax
\expandafter\let\csname endequation*\endcsname\relax
\usepackage{mhchem}
\usepackage{siunitx}
\usepackage{amsmath}
\makeatletter
\newcommand\kv[2]{%
  \ifnum\pdf@strcmp{\unexpanded{#1}}{V}=0 %
     \expandafter\@firstoftwo
  \else
    \expandafter\@secondoftwo
  \fi
    {\textit{#1}\sub{#2}}
    {#1\sub{#2}}%
}
\makeatother
\makeatletter
\newcommand\kvc[3]{%
  \ifnum\pdf@strcmp{\unexpanded{#1}}{V}=0 %
     \expandafter\@firstoftwo
  \else
    \expandafter\@secondoftwo
  \fi
    {\textit{#1}\sub{#2}\super{#3}\negmedspace}
    {#1\sub{#2}\super{#3}\negmedspace}
}
\makeatother
\usepackage{graphicx}
\usepackage{url}
\usepackage{hyperref}
\usepackage{xfrac}
\usepackage[capitalise]{cleveref}
\usepackage{longtable}
\usepackage{algorithm2e}  % for algorithms
\makeatletter
\renewcommand{\@algocf@capt@plain}{above}% formerly {bottom}
\makeatother
\usepackage{tikz}  % for dashed box in algorithm
\crefname{algocf}{Alg.}{Algs.}  % for creffing algorithms
\Crefname{algocf}{Algorithm}{Algorithms}  % for creffing algorithms
\mciteErrorOnUnknownfalse  % ignore error message about SI cites
\begin{document}

\title[Identifying Split Vacancy Defects with Foundation Models and Electrostatics]{Identifying Split Vacancy Defects with Machine-Learned Foundation Models and Electrostatics}

\author{Seán R. Kavanagh$^*$}

\address{Harvard University Center for the Environment, Cambridge, Massachusetts 02138, United States}
\ead{skavanagh@seas.harvard.edu}
\vspace{10pt}

% Despite key roles in material properties, their characterisation is made difficult by their low concentrations.
% Theoretical methods thus dominate defect characterisation research.
\begin{abstract}
Point defects are ubiquitous in solid-state compounds, dictating many functional properties such as conductivity, catalytic activity and carrier recombination.
Over the past decade, the prevalence of metastable defect geometries and their importance to relevant properties has been increasingly recognised.
A striking example is split vacancies, where an isolated atomic vacancy transforms to a stoichiometry-conserving complex of two vacancies and an interstitial (\kv{V}{X} $\rightarrow$ [\kv{V}{X} + \kv{X}{i} + \kv{V}{X}]), which can be accompanied by a dramatic energy lowering and change in behaviour.
These species are particularly challenging to identify from computation, due to the `non-local' nature of this reconstruction.
Here, I present an approach for the efficient identification of these defects, through tiered screening which combines geometric analysis, electrostatic energies and foundation machine learning (ML) models.
This approach allows the screening of all solid-state compounds in the Materials Project database (including all entries in the ICSD, along with several thousand predicted metastable materials), identifying thousands of low energy split vacancy configurations, hitherto unknown. % anything else to mention here?
This study highlights both the potential utility of (foundation) machine-learning potentials, with important caveats, the significant prevalence of split vacancy defects in inorganic solids, and the importance of global optimisation approaches for defect modelling.
%for correctly identifying stable defect geometries.
\end{abstract}

% For two-column output uncomment the next line and choose [10pt] rather than [12pt] in the \documentclass declaration
%\ioptwocol
%
\setcounter{footnote}{0}  % for footnote formatting
\makeatletter
\long\def\@makefntext#1{\parindent 1em\noindent 
 \makebox[1em][l]{\footnotesize\rm$\m@th{[\arabic{footnote}]}$}%
 \footnotesize\rm #1}
\def\@makefnmark{\textsuperscript{[\hbox{${\arabic{footnote}}\m@th$}]}}
\def\@thefnmark{\arabic{footnote}}
\makeatother

% \begin{bibunit}  % section off these references for main text
\fontsize{11}{12}\selectfont
\section*{Introduction}
% Intro would be defects are important, growing appreciation of metastabilities (in general) in recent years, importance of global optimization for defects, a particularly noticeable examples of this is the split vacancy in gallium oxide from Joel, ShakeNBreak despite its power does not find this because it is a non-local reconstruction and other methods don't find it unless using a very large radius in which to generate random structures which would just be prohibitively expensive in the first place. Ideally in the future we will have efficient machine learning models which will be able to perform such exhaustive structure searching, but for now we need to be more intelligent. 
Point defects are an unavoidable feature of bulk solid-state materials, due to the large configurational entropy gain associated with their formation.\cite{freysoldt_first-principles_2014,mosquera-lois_imperfections_2023} % and Squires perspective if preprinted
These defects dictate functional properties and performance in many compounds and applications, from beneficial impacts in semiconductor doping, catalytic activity, ionic conductivity and single-photon emission, to detrimental effects on carrier recombination, pernicious absorption and chemical degradation, to name a handful.\cite{oba_design_2018,asebiah_defect-limited_2024,turiansky_nonrad_2021}
The characterisation and manipulation of defects in materials is thus a primary route to advancing a wide range of technological capabilities, particularly in the realm of energy materials such as solar photovoltaics, transparent conducting materials, batteries and (photo-)catalysts.\cite{kavanagh_intrinsic_2025,zhang_origin_2022,kumagai_alkali_2023}
In recent years, there has been renewed interest in the complexity of defect energy landscapes, with the potential for multiple different locally-stable configurations which contribute to the overall behaviour of the species.\cite{mosquera-lois_search_2021,kononov_identifying_2023,mosquera-lois_identifying_2023}
This can arise when the defect can adopt multiple different bonding configurations, spin states or others.\\  % could cite more here

One of the first well-known examples of such metastability for defects was the case of the so-called `DX-centres' in (Al)GaAs and Si, studied in the 1970s.
Here, puzzling observations of transient behaviour, persistent photoconductivity, charge compensation and large Stokes shifts led researchers to propose that donor defects (D) were forming complexes with an unknown acceptor defect (X) to compensate their charge.\cite{lang_large-lattice-relaxation_1977}
With the help of theoretical studies,\cite{chadi_energetics_1989} it was later revealed that no other defect was involved in these processes, and rather the transformation was driven by the donor defect D displacing significantly off-site to a new bonding configuration, where a negative (acceptor) charge state was then stabilised.
Crucially, there is a small energy barrier to this transition, and so a bias or thermal energy is required to observe this behaviour in either experiment or computation.
Such metastability for point defects has since been shown to impact electron-hole recombination rates in LEDs\cite{alkauskas_role_2016} and photovoltaic devices,\cite{kavanagh_impact_2022,dou_chemical_2023,huang_metastability_2023} oxidation and decomposition in battery cathodes,\cite{squires_oxygen_2024,cen_cation_2023} catalytic activity in oxides,\cite{kehoe_role_2011} charge compensation in chalcogenides,\cite{wang_four-electron_2023} and absorption spectra in II-VI compounds,\cite{lany_metal-dimer_2004} to name a few. \\

\begin{figure}[h]
\centering
\includegraphics[width=\textwidth]{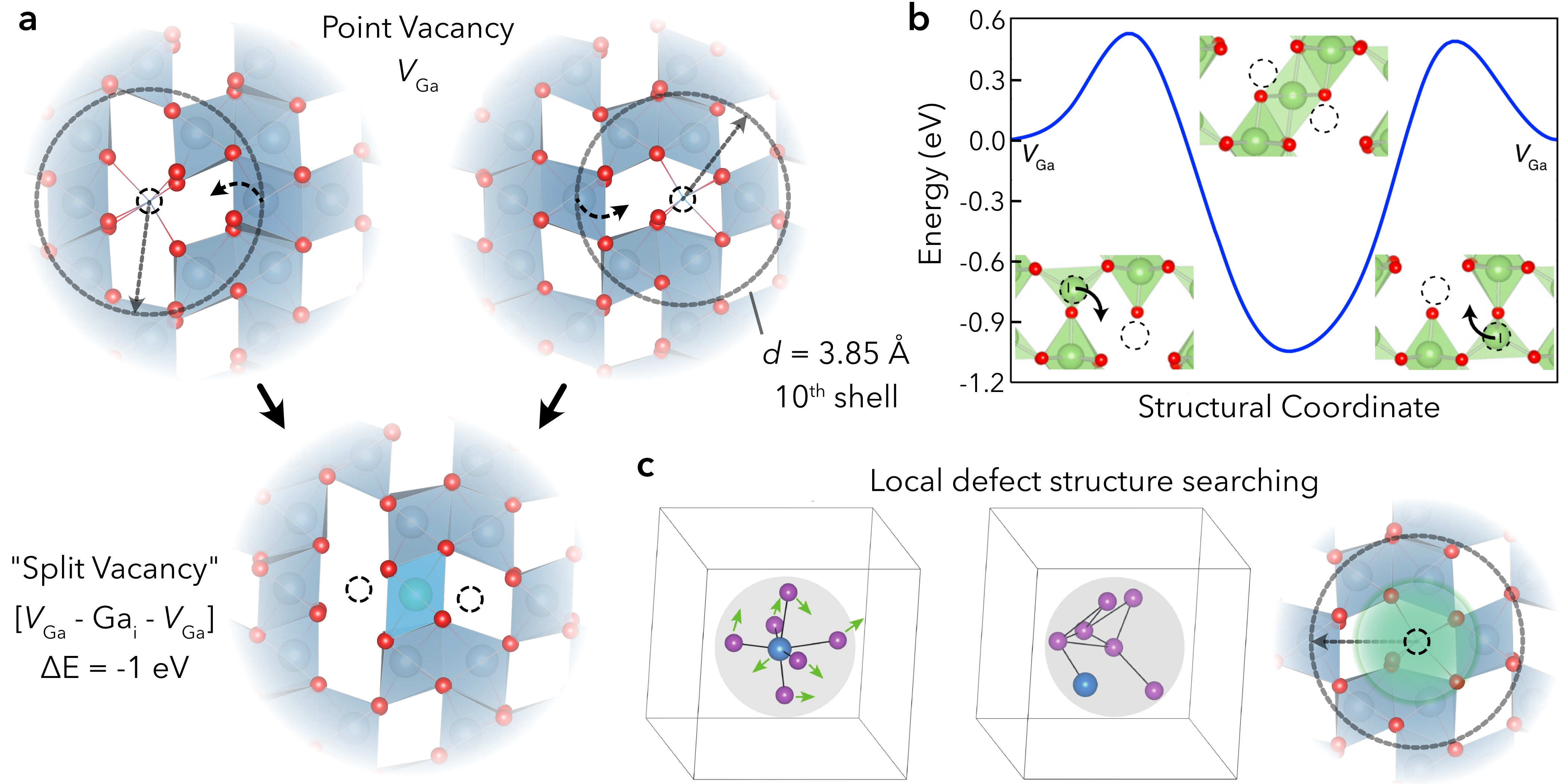}
\caption{Split vacancy configurations in solids. \textbf{(a)} Schematic illustration of the transformation from a single atomic vacancy (top) to a split vacancy geometry (bottom) using \kv{V}{Ga} in $R\bar{3}c$ $\alpha$-\ce{Ga2O3} as an example. Vacancy positions are indicated by the hollow circles, curved arrows depict the movement of the neighbouring cation in transforming from the single vacancy to the split vacancy, and dashed grey circles depict the 10th neighbour shell of the vacancy site ($d$ = \SI{3.85}{\angstrom}). In the bottom image, the interstitial cation within the split vacancy ([\kv{V}{X} + \textbf{\kv{X}{i}} + \kv{V}{X}]) is highlighted in lighter blue.
\textbf{(b)} Potential energy surface (PES) of the tetrahedral-site Ga vacancy in $C2/m$ $\beta$-\ce{Ga2O3}, along the symmetric path from the single vacancy (endpoints) to the split vacancy (middle), adapted with permission from \citeauthor{varley_hydrogenated_2011}\cite{varley_hydrogenated_2011}
\textbf{(c)} Structure searching methods employed for point defects, comprising targeted\cite{mosquera-lois_shakenbreak_2022} or random\cite{huang_dasp_2022,morris_hydrogensilicon_2008,arrigoni_evolutionary_2021,kumagai_insights_2021} local bond distortions (left) and/or chemical identity permutations (centre). Adapted with permission from \citeauthor{huang_dasp_2022}\cite{huang_dasp_2022}. Typical search radii for defect reconstructions are depicted by the shaded green area for \kv{V}{Ga} $\alpha$-\ce{Ga2O3} on the right, significantly smaller than the \kv{V}{X}-\kv{V}{X} distance in [\kv{V}{X} + \kv{X}{i} + \kv{V}{X}] split-vacancy complexes. 
}
\label{fig:Split_Vacancies_Intro_Fig}
\end{figure}

A particularly striking case of such large defect reconstructions is that of `split vacancies' where, starting from the simple vacancy picture, a nearby atom displaces toward an interstitial site adjacent to the vacancy, effectively creating an additional vacancy and interstitial in the process; \cref{fig:Split_Vacancies_Intro_Fig}a.
A split vacancy can thus be thought of as a stoichiometry-conserving complex of two vacancies and an interstitial (\kv{V}{X} $\rightarrow$ \kv{X}{i} + 2\kv{V}{X}).
A dramatic lowering of the defect energy and change in behaviour can accompany this large structural transformation.
For instance, one of the most well-known cases of split vacancies is that of \kv{V}{Ga} in \ce{Ga2O3}, which has been extensively studied as a promising material for power electronics and transparent conducting oxides.\cite{frodason_multistability_2021,portoff_hydrogen_2023}
As reported by \citeauthor{varley_hydrogenated_2011}\cite{varley_hydrogenated_2011} and shown in \cref{fig:Split_Vacancies_Intro_Fig}b, negatively-charged Ga vacancies --- the dominant acceptor species --- can form split-vacancy complexes which lower their energy by $\sim$\SI{1}{eV}.\footnote{~This behaviour was originally reported for monoclinic $C2/m$ $\beta$-\ce{Ga2O3}, but has since been observed for the corundum-structured $R\bar{3}c$ $\alpha$ and orthorhombic $\kappa$ phases as well.\cite{fowler_metastable_2024,lei_linking_2015,mazzolini_engineering_2024,okumura_mocvd_2024}}
This transformation has several crucial implications for the electronic and defect behaviour of \ce{Ga2O3}.
Firstly, the large energy lowering of the negatively-charged cation vacancy (\kvc{V}{Ga}{-3}) greatly increases its concentration and makes it a shallower acceptor species.
Being the most favourable intrinsic acceptor in \ce{Ga2O3} -- which is typically doped $n$-type -- this places greater limitations on electron dopability by reducing the electron doping window and enhancing ionic charge compensation.
% In addition to greatly increasing their concentration and making them a much shallower acceptor species, due to the large energy lowering
Moreover, the split vacancy geometry of \kv{V}{Ga} is found to be key to ion migration pathways,\cite{frodason_migration_2023} which are relevant for the diffusion of technologically-important dopants and impurities in \ce{Ga2O3}.
These split configurations of Ga vacancies in \ce{Ga2O3} have since been verified by a number of experimental measurements, including positron annihilation spectroscopy,\cite{karjalainen_split_2020,karjalainen_split_2021} electron paramagnetic resonance (EPR),\cite{skachkov_computational_2019,von_bardeleben_proton_2019} scanning transmission electron microscopy (STEM),\cite{johnson_unusual_2019} and vibrational spectroscopy with hydrogenated samples.\cite{qin_editors_2019,stavola_tutorial_2024,weiser_structure_2018}\\

\AtBeginEnvironment{figure}{\mathindent=0pt}  % don't indent figure caption, default in IOP format

% \textbf{In2O3 is really important? Should look at / note this?}
Such split vacancies have since been shown to exist in a small handful of other structurally-related and technologically-relevant compounds, such as the $R\bar{3}c$ corundum-structured polymorph $\alpha$-\ce{Ga2O3}, along with corundum-like compounds; \ce{Al2O3}, \ce{In2O3}, \ce{Ca3N2} and \ce{Mg3N2}.\cite{fowler_metastable_2024,lei_linking_2015,kononov_identifying_2023}
Recently, \citeauthor{fowler_metastable_2024}\cite{fowler_metastable_2024} discussed the known cases of these split vacancies; including those mentioned above, beta-tridymite \ce{SiO2} and a handful of metastable split vacancies in some rutile compounds and \ce{Cu2O}. 
In an investigation of \ce{Sb2O5} as a candidate transparent conducting oxide earlier this year, \citeauthor{li_computational_2024}\cite{li_computational_2024} reported a split-vacancy structure for \kv{V}{Sb} found using the \texttt{ShakeNBreak}\cite{mosquera-lois_shakenbreak_2022,mosquera-lois_identifying_2023} approach, which lowers the vacancy energy by over \SI{2}{eV}.
While this small set of known split vacancy geometries is mostly comprised of \emph{cation} vacancy defects, split vacancy structures have also been observed for some \emph{anion} vacancies, such as \kv{V}{O} in \ce{TiO2}\cite{arrigoni_evolutionary_2021} \& \ce{Ba2TiO4}\cite{kumagai_insights_2021} and \kv{V}{N} in \ce{Mg3N2} \& \ce{Ca3N2}\cite{fowler_metastable_2024}.\\

% The typically dilute concentration of defects, despite their major impact on macroscopic properties, renders experimental characterisation extremely challenging. 
% Thus, theoretical methods represent the primary avenue for the investigation of point defects at the atomic-scale, with computational predictions of defect behaviour (and resultant impact) often being compared to experimental measurements of downstream global properties, such as carrier concentrations, solar cell efficiency, ionic conductivity or catalytic activity.
Theoretical methods often represent the primary avenue for the investigation of point defects at the atomic scale, due to an inherent difficulty in experimentally characterising dilute localised species.
% Computational predictions of defect behaviour (and resultant impact) are often then compared to experimental measurements of downstream global properties, such as carrier concentrations, solar cell efficiency, ionic conductivity or catalytic activity.
% The accuracy of such computational approaches, when employing appropriate levels of theory and careful methodology, has been well verified through comparisons with experiment.\cite{freysoldt_first-principles_2014,kavanagh_rapid_2021,stavola_tutorial_2024,mosquera-lois_imperfections_2023}
Metastability at defects presents a challenge to computational methods, however.
Defect modelling involves the simulation of a defect embedded in the bulk compound (typically using a large periodic supercell), from which the formation energy and related properties can be computed.
This requires some initialisation of the defect state in terms of geometry and spin, before relaxing to the local minimum energy arrangement -- typically via gradient descent.
However, when defects exhibit multiple locally-stable minima, this single predicted arrangement will not give the full picture of defect behaviour. % shown to be the case for many defects across diverse material classes\cite{mosquera-lois_identifying_2023}
In many cases, this state will be a higher-energy metastable configuration, which can lead to inaccurate predictions of defect properties, such as defect and charge-carrier concentrations, recombination activity, diffusion barriers and more.
Indeed, as implied by the potential energy surface (PES) in \cref{fig:Split_Vacancies_Intro_Fig}b, the identification of the split vacancy \kv{V}{Ga} in $\beta$-\ce{Ga2O3} was a serendipitous discovery by \citeauthor{varley_hydrogenated_2011},\cite{varley_hydrogenated_2011} where a Nudged Elastic Band (NEB) calculation of the (simple) vacancy migration pathway revealed that the expected transition state (the split vacancy) was in fact the ground-state arrangement.
A handful of approaches have been proposed to counteract this issue and target a global optimisation strategy for defects.\cite{arrigoni_evolutionary_2021,huang_dasp_2022,mosquera-lois_search_2021,morris_hydrogensilicon_2008}
For instance, within the \texttt{ShakeNBreak}\cite{mosquera-lois_shakenbreak_2022,mosquera-lois_identifying_2023} approach, a set of candidate geometries are generated by distorting the local bonding environment of the defect according to some simple chemical guiding principles, along with constrained random displacements (`rattling') to break symmetry and disrupt the long-range lattice potential, in order to coarsely sample various regions of the defect energy landscape. % of all atoms in the structure 
Despite its relative simplicity and computational efficiency, this approach has been found to perform surprisingly well in identifying structural reconstructions and metastabilities at defects, and has demonstrated the prevalence and importance of defect reconstructions across diverse materials classes.\cite{mosquera-lois_identifying_2023} \\ % could cite more here...
% The wide prevalence of such behaviour was recently demonstrated,\cite{mosquera-lois_identifying_2023}\textbf{(again could be earlier?)} using the ShakeNBreak approach.
% \textbf{Do we note significantly the importance of this here? Energy lowering has important implications in these specific examples, and of course in general}

Split vacancies present a distinct challenge to current defect structure-searching approaches however, with most failing to identify ground-states split vacancy geometries in the majority of cases.
Indeed, both in this work and in unpublished work from Dr Joel Varley, it was confirmed that \texttt{ShakeNBreak} fails to identify the split-vacancy groundstate for \kv{V}{Ga} in \ce{Ga2O3}, when starting from the isolated vacancy geometry.
While \texttt{ShakeNBreak} did manage to identify the \SI{2}{eV} lower-energy split-vacancy ground-state for \kv{V}{Sb} in \ce{Sb2O5} --- when using large bond distortions and atom rattling --- this is deemed to be a rare case where the lower symmetry and reduced cation coordination give a lower energy barrier to the transformation and allows it to be identified with semi-local structure searching.
This can be attributed to the built-in locality of these optimisation approaches, which attempt to leverage our physical intuition regarding defect behaviour to bias the search space and boost computational efficiency.
Namely, these approaches make use of the `molecule-in-a-solid' nature of defects, where interactions are inherently short-range and dominated by the first and next-nearest neighbours, with most (known) defect reconstructions involving some perturbation of this highly-\emph{local} bonding environment.
As such, these approaches often start from the unperturbed defect structure (e.g. the simple removal of an atom to create a vacancy, or mutation of the chemical identity to create a substitution), and then apply \emph{local} geometry perturbations to efficiently scan accessible reconstructions, as depicted in \cref{fig:Split_Vacancies_Intro_Fig}c.\\

The transformation of isolated vacancies to split vacancies; \kv{V}{X} $\rightarrow$ \kv{X}{i} + 2\kv{V}{X}, is a \emph{non-local} process however, as it involves the movement of a host atom which is initially quite far from the defect site (\SI{3.85}{\angstrom} for \kv{V}{Ga} in $\alpha$-\ce{Ga2O3} -- \cref{fig:Split_Vacancies_Intro_Fig}a, corresponding to the 10th nearest-neighbour shell or the 28th-closest atom) to a position much closer to the original vacancy site --- around the midpoint of the original separation.
This point is further verified by structural analysis of split vacancies identified in this work, shown in \cref{sifig:displacements}, where the largest atomic displacements relative to bulk positions occur for atoms located 3-\SI{5}{\angstrom} away from the single vacancy site for split vacancies, as opposed to 1-\SI{2}{\angstrom} for simple point vacancies.
Consequently, `local' structure-searching methods such as \texttt{ShakeNBreak} which target distortions involving the first few neighbour shells are expected to struggle at identifying these `non-local' reconstructions. \\

In this work, I set out to investigate the prevalence of split vacancies in solid-state compounds. 
Currently only a small handful of these species, mentioned above, are known.
Is this the result of an inherent rarity in nature, or simply because we have not had the computational tools to efficiently search for these species (or both)?
% -- as was the case of energy-lowering reconstructions at defects in general until recently --
Indeed the prevalence of defect metastability in general is more appreciated due to recent efforts for efficiently identifying this behaviour. % \cite{arrigoni_evolutionary_2021,huang_dasp_2022,mosquera-lois_shakenbreak_2022,mosquera-lois_identifying_2023,mosquera-lois_search_2021,morris_hydrogensilicon_2008,lee_investigation_2024}
As such, I begin by analysing the energetic driving factors for split-vacancy defect formation.
Leveraging these insights, I develop an efficient approach for their identification using geometric and electrostatic analyses, and use it to screen for split cation vacancies in several hundred metal oxides.
I then demonstrate that machine-learned interatomic potentials are effective tools for accelerating these defect potential energy surface evaluations, which allows the screening of all compounds in the Materials Project database (which includes all entries in the Inorganic Crystal Structure Database (ICSD), along with several thousand computationally-predicted materials), identifying thousands of hitherto unknown low energy split vacancy defects.
Finally, I conclude with a discussion of the wider implications of these findings, the generality of this approach for (defect) energy surface exploration, and the potential use of machine-learned potentials for defect modelling.

\section*{Methods}
\begin{figure}[h]
\centering
\includegraphics[width=\textwidth]{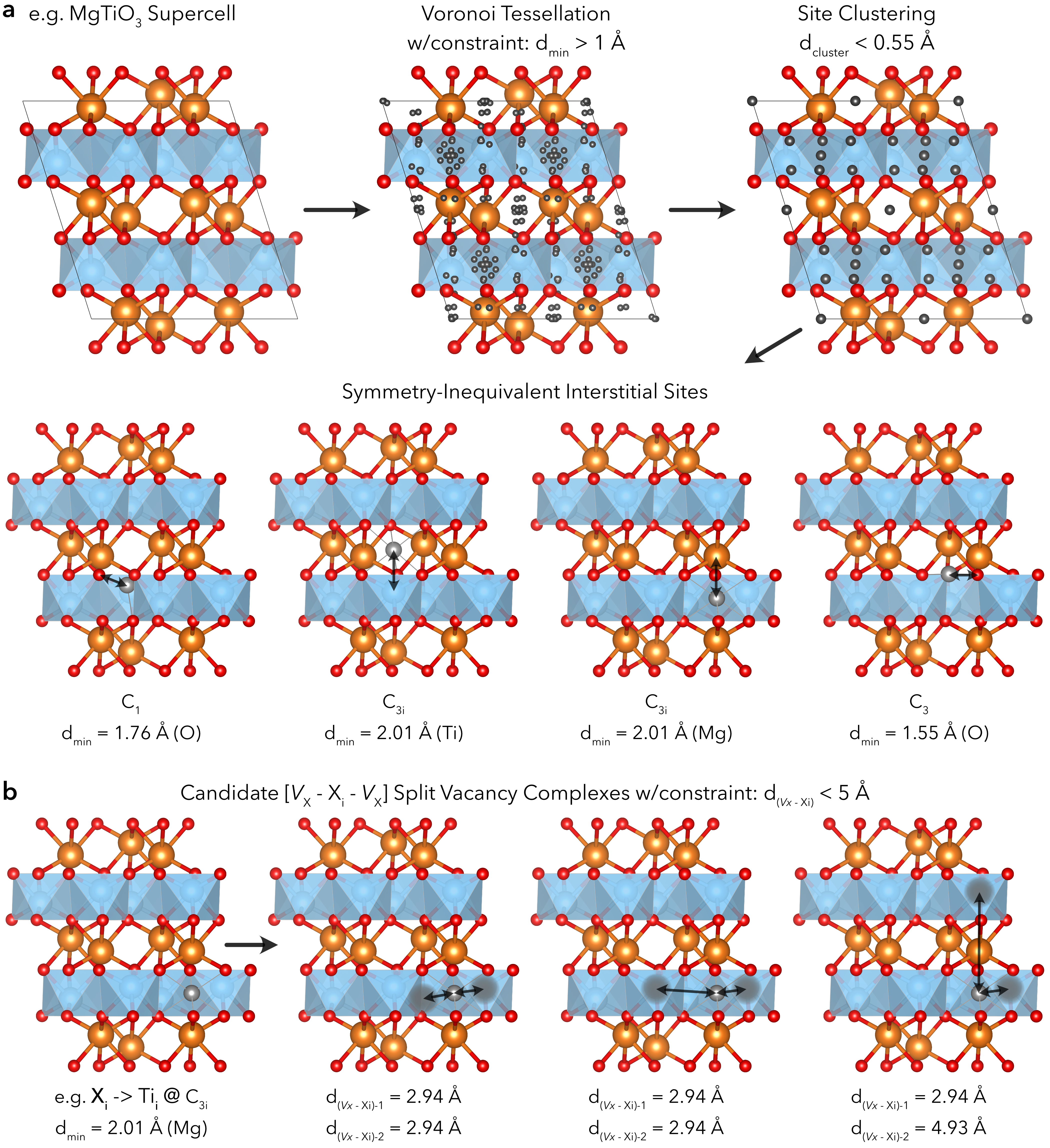}
\caption{Generation of interstitial and split vacancy geometries via \texttt{doped} 
\textbf{(a)} Interstitial generation workflow, using \ce{MgTiO3} as an example. Mg cations are in orange, Ti in blue and O in red. Candidate interstitial sites are shown as grey spheres. The point symmetries and minimum distances to host atoms of the final symmetry-inequivalent interstitial sites are given underneath.
\textbf{(b)} Split vacancy generation workflow, using the titanium interstitial (\kv{Ti}{i}) at the $C_{3i}$ site (with Mg nearest neighbour) in \ce{MgTiO3} as an example. Vacancies are indicated by faded grey circles, and vacancy-interstitial (\kv{V}{X}-\kv{X}{i}) distances are listed underneath and indicated by double-headed black arrows.
}
\label{fig:Interstitial_and_Split_Vacancy_Generation_Methods}
\end{figure}

The \texttt{doped}\cite{kavanagh_doped_2024} defect simulation package was used extensively in this work, for the generation of symmetry-inequivalent defect geometries, supercell generation, geometric (distance) and symmetry analysis, input file generation, utility functions for serialization and more.
As illustrated in \cref{fig:Interstitial_and_Split_Vacancy_Generation_Methods}a, candidate interstitial sites were generated using Voronoi tessellation, where only sites with a minimum distance of \SI{1.0}{\angstrom} from the host atomic framework were retained (\texttt{min\_dist} in \texttt{doped}\cite{kavanagh_doped_2024}). 
Interstitial sites separated by distances less than the clustering tolerance (\texttt{clustering\_tol}; set to \SI{0.55}{\angstrom} here) were combined as one candidate site, with preference for the higher symmetry site conditioned on the minimum distances to host atoms, controlled by \texttt{symmetry\_preference}.
As illustrated in \cref{fig:Interstitial_and_Split_Vacancy_Generation_Methods}b, candidate split vacancy geometries were generated by taking each interstitial \kv{X}{i} and enumerating all possible symmetry-inequivalent vacancy-interstitial-vacancy (\kv{V}{X}-\kv{X}{i}-\kv{V}{X}) complexes with \kv{V}{X}-\kv{X}{i} distances less than \SI{5}{\angstrom}.
The full screening workflow employed in this work is summarised in \cref{alg}.
A distance-based classification algorithm was implemented in \texttt{doped} to classify simple, split and non-trivial vacancy geometries, detailed in \cref{SI:Split_Vacancy_Classification}.
In all cases, relative energies are taken as the simple difference between Ewald Summation energies, for the electrostatic model, or supercell energies for Density Functional Theory (DFT) calculations using the same supercell size and charge state, without finite-size corrections.
For the Ewald Summation electrostatic model, the dependence of candidate split vacancy relative energies on supercell size was tested, and found not to significantly affect energy differences for the supercell sizes used here (minimum periodic image distances of \SI{10}{\angstrom}), while the impact of finite-size corrections on relative DFT energies is discussed in \cref{SI:Finite_Size_Corrections}.
\\

All DFT calculations were performed within periodic boundary conditions through the Vienna Ab Initio Simulation Package (VASP).\cite{kresse_ab_1993,kresse_efficiency_1996,kresse_efficient_1996,kresse_ab_1994,kresse_ultrasoft_1999,gajdos_linear_2006} 
Using the projector-augmented wave (PAW) method, scalar-relativistic pseudopotentials were employed to describe the interaction between core and valence electrons.\cite{blochl_projector_1994}
% For the Materials Project database... PBE\cite{perdew_generalized_1996}
For hybrid DFT calculations, the PBE0 hybrid functional was used.\cite{adamo_toward_1999}
For the initial test cases of known split vacancies (\cref{table:split_vacancy_energies}) and split cation vacancies in metal oxides (with host compounds taken from the database of \citeauthor{kumagai_insights_2021}\cite{kumagai_insights_2021}), the PBEsol\cite{perdew_restoring_2008} GGA DFT functional was used for the semi-local DFT calculations.
Calculation parameters and defect supercells were set to be consistent with that used by \citeauthor{kumagai_insights_2021}\cite{kumagai_insights_2021} (using the \texttt{vise} package), avoiding the need for re-relaxation of the bulk lattice parameters, while also using some of the default relaxation settings in \texttt{ShakeNBreak}\cite{mosquera-lois_shakenbreak_2022} (real-space projections, energy convergence thresholds, force minimisation algorithm etc) which have been well-tested for their accuracy and efficiency for defect structure-searching.\cite{mosquera-lois_identifying_2023}  % Squires perspective if possible
Here, this corresponded to the use of a plane-wave energy cutoff of \SI{400}{eV} and $\Gamma$-point only sampling for the defect supercells (which ranged from 60 to 500 atoms in size; \cref{sifig:Supercell_Size_and_Length_Distributions}), and a maximum atomic force convergence threshold of \SI{0.01}{eV/\angstrom}.
When screening compounds from the Materials Project database,\cite{jain_materials_2013} supercells were generated using the \texttt{doped}\cite{kavanagh_doped_2024} algorithm which scans all supercell expansions of the primitive unit cell (including non-diagonal matrices), and selects that with the lowest number of atoms which satisfies the minimum image distance and atom count constraints --- set to \SI{10}{\angstrom} and 50 atoms respectively.
As shown in \cref{alg}, supercells were generated at the start of the workflow for each host compound, with the same supercell size used for electrostatic, machine-learned (ML) model and DFT energy evaluations.
For the stable nitrides test set, the PBE\cite{perdew_generalized_1996} GGA DFT functional was used for DFT relaxations to match the Materials Project\cite{jain_materials_2013} computational setup, combining the \texttt{MPRelaxSet}\cite{ong_python_2013} parameters (e.g. \SI{520}{eV} energy cutoff) with the default \texttt{ShakeNBreak}\cite{mosquera-lois_shakenbreak_2022} relaxation settings.
The \texttt{MACE-mp} foundation model\cite{batatia_foundation_2024} was used as the primary ML interatomic potential for ML-accelerated screening in this work, for which a number of tests of speed and accuracy for model size, optimiser algorithm and float precision were performed, with results and discussion provided in \cref{SI:MACE_tests}.
The \texttt{nequip}\cite{batzner_e3-equivariant_2022,tan_high-performance_2025} ML potential architecture was also trialled, achieving similar accuracies.
\\

Random displacements of atomic positions (`rattling') to break symmetry can aid the identification of lower-energy, lower-symmetry geometries by gradient optimisers,\cite{kumagai_insights_2021,mosquera-lois_shakenbreak_2022,krajewska_enhanced_2021} however this did not significantly increase the number of lower energy split vacancies identified in this study, which is attributed to the fact that split vacancies mostly have high symmetries (\cref{fig:MP_Screening}b)\cite{fowler_metastable_2024} and fully-ionised defects are much less likely to break symmetry.\cite{mosquera-lois_identifying_2023,mosquera-lois_machine-learning_2024,lany_anion_2005}
However, rattling did reveal some bulk phase transformations to lower-symmetry, lower-energy compounds (\href{https://shakenbreak.readthedocs.io/en/latest/Tips.html#bulk-phase-transformations}{shakenbreak.readthedocs.io/en/latest/Tips.html\#bulk-phase-transformations}),\cite{neilson_oxygen_2024,krajewska_enhanced_2021} which have imaginary phonon modes off the $\Gamma$ point.
These included \ce{Ta2O5} ($Pmmn \rightarrow P2_1/m$; $\Delta E$ = \SI{22}{meV/fu}), \ce{TiTl2(GeO3)3} ($P6_3/m \rightarrow P_1$; $\Delta E$ = \SI{48}{meV/fu}) and \ce{Li5BiO5} ($C2/m \rightarrow P2_1$; $\Delta E$ = \SI{22}{meV/fu}).
In some cases, such as for the photoconductive \ce{KTaO3} system,\cite{santillan_room-temperature_2024} the symmetry perturbation introduced by (initial) split vacancy geometries resulted in the identification of lower energy \emph{point} vacancy structures, similar to the effect of rattling.
\\
% This was also recently found for UO$_{2\pm x}$ using \texttt{ShakeNBreak}\cite{neilson_oxygen_2024} and \ce{Cs2AgBiBr6}, and \ce{Cs3Bi2Br9}\cite{krajewska_enhanced_2021}\\

The screening approach employed here would not have been possible without several of the efficient algorithms in \texttt{doped};\cite{kavanagh_doped_2024} including fast geometric and symmetry analysis of complex structures (to determine symmetry-inequivalent defect sites and complexes), oxidation and charge state estimation, flexible generation parameters to maximise efficiency and reduce memory demands when screening thousands of complex compounds, and more. % any more to mention?
The stable generation of intrinsic and complex defects with reasonable estimated charge states in all compounds on the Materials Project,\cite{jain_materials_2013} including low-symmetry multinary compositions with large unit cells ($>$ 200 atoms) such as \ce{Na30Mg4Ta20Si33(SO48)3} and \ce{Na7Zr2Si5Ge2PO24}, was a powerful test of robustness.
Some additional methodological details are given in \cref{SI:Additional_Methods}.
% Any relevant notes from notebooks?

\section*{Results}
\subsection*{Factors Driving Split Vacancy Formation}
The identification of split vacancy structures is not, at first glance, trivial.
Local structure-searching approaches, despite their general success, do not succeed in identifying these species in most known cases.
A brute force enumeration approach, where each possible vacancy-interstitial-vacancy combination (i.e. split vacancy) is trialled with a coarse total energy calculation -- DFT or otherwise -- is also out of the question.
For instance, if we enumerate all possible symmetry-inequivalent \kv{V}{X}-\kv{X}{i}-\kv{V}{X} complexes with \kv{V}{X}-\kv{X}{i} distances less than \SI{5}{\angstrom}, with interstitial sites determined by Voronoi tessellation,\cite{kavanagh_doped_2024,kononov_identifying_2023} this gives 540 combinations in $\alpha$-\ce{Ga2O3}, 1267 in $\beta$-\ce{Ga2O3}, 452 in \ce{Sb2O5} and even larger numbers in compounds with lower symmetry and/or greater multinarity. % 1004 for Cu2O, 26,000 for H in CdH20N6OF2
In most cases, it is infeasible to perform DFT supercell calculations for such a large number of trial structures, especially when considering each symmetry-inequivalent vacancy site and charge state. \\

\begin{figure}[h]
\centering
\includegraphics[width=\textwidth]{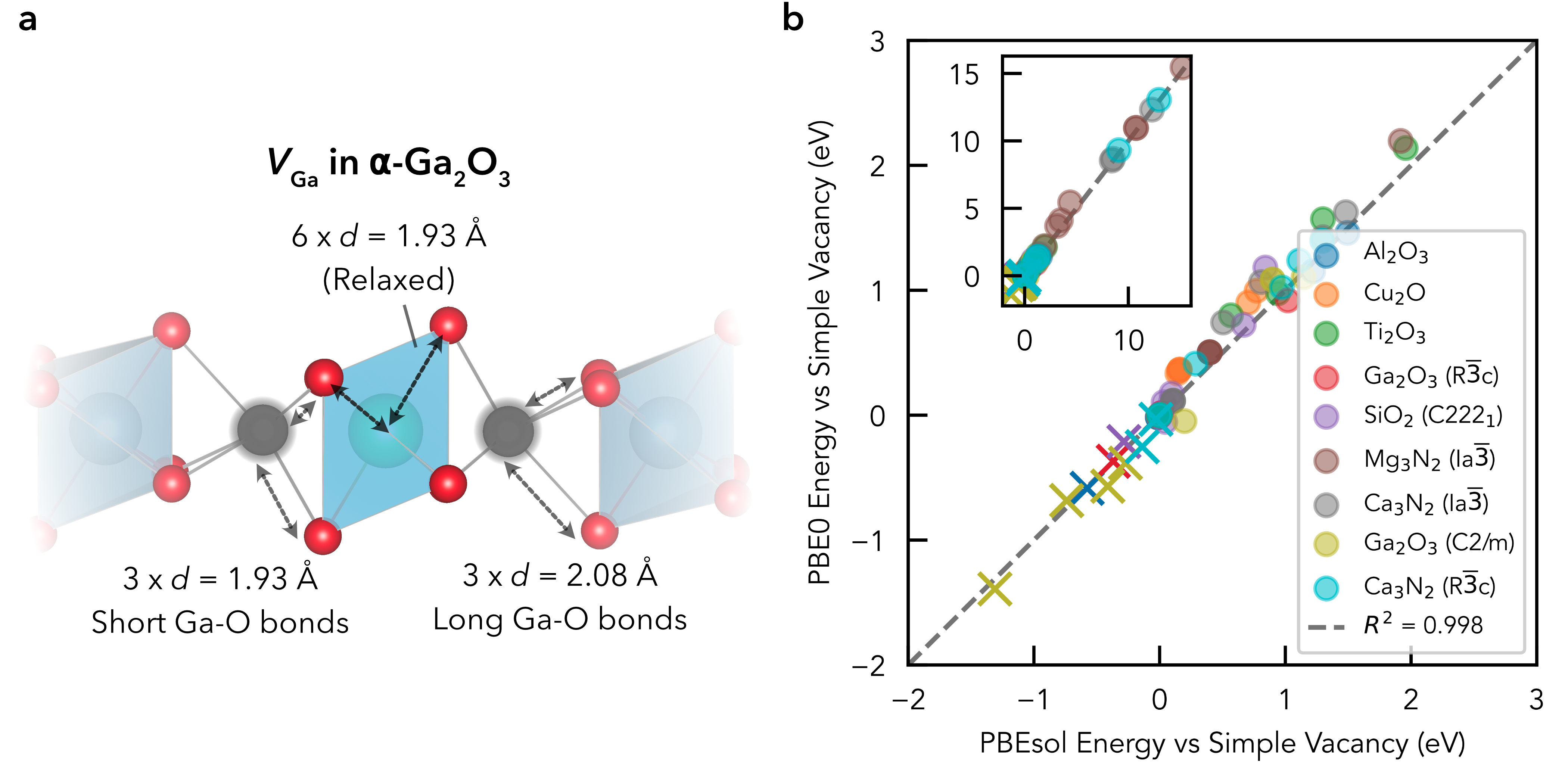}
\caption{Geometric and DFT energy analysis of split vacancies. 
\textbf{(a)} Geometric analysis of the split vacancy for \kv{V}{Ga} in $R\bar{3}c$ $\alpha$-\ce{Ga2O3}, indicating short and long cation-anion bond lengths. The interstitial cation within the split vacancy ([\kv{V}{X} + \textbf{\kv{X}{i}} + \kv{V}{X}]) is highlighted in lighter blue as in \cref{fig:Split_Vacancies_Intro_Fig}, and vacancy positions are indicated by the semi-transparent grey circles.
\textbf{(b)} Energies of relaxed candidate split vacancy configurations relative to the lowest energy symmetry-inequivalent simple vacancy, using semi-local GGA (PBEsol; $x$-axis) and hybrid DFT (PBE0; $y$-axis). Split vacancies which are lower energy than the lowest energy simple vacancy are denoted by $\times$ symbols, and the same plot over a wider energy range is shown inset.
This dataset includes cation vacancies for oxides and the anion vacancy for nitrides (anti-corundum-like structures).
}
\label{fig:Geometric_and_PBE0}
\end{figure}

The efficient identification of split-vacancy clusters will thus require a significant reduction of this search space to a tractable number of candidate structures.
In this regard, it is beneficial to understand the driving forces of their formation in order to leverage these physical insights in our computational strategy.
Here we take \kvc{V}{Ga}{-3} in the $R\bar{3}c$ corundum-structured $\alpha$-\ce{Ga2O3} phases as an example (shown in \cref{fig:Split_Vacancies_Intro_Fig}a) -- for which the same split vacancy transformation is known to occur in the isostructural $\alpha$-\ce{Al2O3} (sapphire).\cite{kononov_identifying_2023}
As shown in \cref{fig:Geometric_and_PBE0}a, each cation octahedron has three short ($\sim$\SI{1.93}{\angstrom}) and three long ($\sim$\SI{2.08}{\angstrom}) Ga-O bond lengths.
In the simple vacancy we thus have three \emph{short} and three \emph{long} cation-anion dangling bonds, while in the split vacancy (\kv{V}{X} $\rightarrow$ \kv{V}{X} + \kv{X}{i} + \kv{V}{X}) we instead have three \emph{long} cation-anion dangling bonds for each of the two vacant octahedra.
The formation energy of a defect is essentially derived from the energetic cost of the bond breaking and creation induced by its formation, and associated strain costs.
Here, the breaking of only the \emph{longer} cation-anion bonds is found to result in a more energetically favourable arrangement than breaking both \emph{short} and \emph{long} cation-anion bonds. % (and creation of the interstitial-cation-anion bonds)
While not entirely straightforward due to the lower symmetry and variable preferences for shorter/longer cation-anion bond lengths, the connection between simple bond counting and the energetic ordering of simple vs split vacancies here suggests that their formation can be estimated through analysis of the host crystal structure.
In particular, the existence of the low energy split (cation) vacancy evidently requires an interstitial position located adjacent to two cation sites, which has similar (or greater) anion coordination compared to the host cation site.\\
% \footnote{~Note that while the discussion here focuses on split \emph{cation} vacancies, the same considerations hold for anion vacancies} – tbf, we mention split anion vacancies in the intro, and are hoping to add at least one of these to the table below, so maybe not necessary to highlight here as the discussion below is mostly pretty general

\AtBeginEnvironment{table}{\mathindent=0pt}  % don't indent figure caption, default in IOP format
\begin{table}[h]
\def\arraystretch{1.25}  % cleaner, more space
\begin{minipage}{\textwidth}
\caption{Energies in electronvolts (eV) of split vacancy configurations relative to the lowest-energy symmetry-inequivalent simple vacancy, in known occurrences.\cite{li_computational_2024,kononov_identifying_2023,fowler_metastable_2024,lei_linking_2015,frodason_multistability_2021}. Relative energies calculated using both the semi-local PBEsol and hybrid non-local PBE0 DFT functionals are given. Asterisks ($^*$) denote metastable split-vacancy configurations, for cases where there are multiple symmetry-inequivalent low-energy split vacancy states. \\
}  % Could add the ES energies here too if we thought useful -- nah will do in an extra figure
\begin{center}
\resizebox{\textwidth}{!}{%
\centering
\begin{tabular}{l|c|c|c|c|c|c|c|c}
\hline
\textbf{Functional} & 
\begin{tabular}[t]{@{}c@{}}\textbf{\ce{Al2O3}} \\ $R\bar{3}c$ \\ \kvc{V}{Al}{-3} \end{tabular} & 
\begin{tabular}[t]{@{}c@{}}\textbf{\ce{Ga2O3}} \\ $R\bar{3}c$ \\ \kvc{V}{Ga}{-3} \end{tabular} & 
\begin{tabular}[t]{@{}c@{}}\textbf{\ce{Ga2O3}} \\ $C2/m$ \\ \kvc{V}{Ga}{-3} \end{tabular} & 
\begin{tabular}[t]{@{}c@{}}\textbf{\ce{Ga2O3}} \\ $C2/m$ \\ \kvc{V}{Ga}{-3}$^*$ \end{tabular} & 
\begin{tabular}[t]{@{}c@{}}\textbf{\ce{Ga2O3}} \\ $C2/m$ \\ \kvc{V}{Ga}{-3}$^{**}$ \end{tabular} & 
\begin{tabular}[t]{@{}c@{}}\textbf{\ce{SiO2}} \\ $C222_1$ \\ \kvc{V}{Si}{-4} \end{tabular} & 
\begin{tabular}[t]{@{}c@{}}\textbf{\ce{Ca3N2}} \\ $R\bar{3}c$ \\ \kvc{V}{N}{+3} \end{tabular} &
\begin{tabular}[t]{@{}c@{}}\textbf{\ce{Sb2O5}}\footnotemark \\ $C2/c$ \\ \kvc{V}{Sb}{-5} \end{tabular}
\\
\hline
PBEsol & -0.58 & -0.37 & -1.31 & -0.74 & -0.41 & -0.28 & -0.14 & -1.43 \\
PBE0   & -0.59 & -0.37 & -1.39 & -0.68 & -0.57 & -0.21 & -0.25 & -1.47 \\
\hline
\end{tabular}
}
\end{center}
\label{table:split_vacancy_energies}
\footnotetext{\scriptsize{~For \ce{Sb2O5}, the listed values correspond to the energy differences between the ground and metastable split vacancy states, as the PBEsol simple vacancy \kvc{V}{Sb}{-5} geometry destabilises during PBE0 relaxation, making the split vs simple vacancy comparison invalid for this case.}}
\end{minipage}
\end{table}
% This is just taking the PBEsol-relaxed structures, same lattice parameters, and continuing with PBE0

Taking a selection of the known cases of (meta)stable split vacancy defects, discussed in the introduction, I calculate the relative energies of the split and simple vacancy configurations using both hybrid DFT (PBE0) and semi-local GGA DFT (PBEsol), and tabulate the results in \cref{table:split_vacancy_energies}.
In most cases, there is good agreement between hybrid and semi-local DFT for the relative energies, between split and simple vacancy configurations, as well as between metastable split vacancy states.
This is further demonstrated in \cref{fig:Geometric_and_PBE0}b, where we see high correlation ($R^2 = 0.998$) between the PBEsol and PBE0 relative energies for these defects.
At first, this may seem surprising, as semi-local DFT is notoriously inaccurate for simulating defects in semiconductors and insulators.
Indeed, we have tested if defect structure-searching with \texttt{ShakeNBreak}\cite{mosquera-lois_shakenbreak_2022} could be performed using semi-local DFT to identify distinct stable defect geometries, before using higher-level theories to relax and compute energies with greater accuracy, but found that it was unable to even \emph{qualitatively} identify ground-state defect geometries in around \SI{50}{\%} of the reconstructions (missed by standard relaxations) reported in \citeauthor{mosquera-lois_identifying_2023}\cite{mosquera-lois_identifying_2023} -- let alone give reasonable estimates of relative energies. % cite Squires perspective here if preprinted
In each case, this error could be attributed to either self-interaction and resulting spurious delocalisation errors inherent to semi-local DFT (inhibiting charge localisation, which often drives these structural reconstructions) or the related band gap underestimation (spuriously destabilising certain defect charge states).\cite{kavanagh_doped_2024,nicolson_cu2sise3_2023,kumagai_insights_2021}  %  unstable and thus delocalising charge to the band edges to give `false charge states'
However, this is not the case for \emph{fully-ionised} defect charge states,\footnote{~Fully-ionised charge states refer to the case where all atoms are assumed to be in their formal oxidation states, and so e.g. in \ce{Al2O3} addition of an \kvc{O}{}{-2} anion to create an interstitial would give \mbox{\kvc{O}{i}{-2}}, substitution of an oxygen site with aluminium would give \kvc{Al}{O}{+5}, removal of an \kvc{Al}{}{+3} cation to create a vacancy would give \kvc{V}{Al}{-3} etc. Fully-ionised charge states are often the most stable charge states for defects in semiconducting and insulating solids.} where there is no excess charge and so no requirement for charge localisation to stabilise the defect.
In these cases, the main contributors to formation energies are typically cohesive energies (i.e. bond breaking energies) and electrostatic effects.
This is additionally why fully-ionised defect species tend to show \emph{less} structural reconstructions than other charge states,\footnote{~Many of the reconstructions we do find for fully-ionised defects are where ionised interstitials move to lower their electrostatic and strain energies, similar to the behaviour of split vacancies here.\cite{wang_upper_2024,mosquera-lois_identifying_2023}} as the lack of excess charge to localise results in less degrees of freedom in the defect geometry energy landscape, with ionic rather than covalent interactions dominating.
The lack of charge localisation and dominance of ionic interactions in fully-ionised defect species means that semi-local DFT often performs adequately for these defects, without the need for improved exchange-correlation descriptions from hybrid DFT.\cite{kumagai_insights_2021,kononov_identifying_2023}
It is important not to misinterpret this point, semi-local DFT will fail miserably in the vast majority of cases beyond fully-ionised defect states, which are usually only a subset of the relevant defects in a given material.
Indeed, this can be seen from the results of \citeauthor{kononov_identifying_2023}\cite{kononov_identifying_2023} for defects in $R\bar{3}c$ $\alpha$-\ce{Al2O3} (a.k.a. sapphire), where the relative energies of various defect configurations in different charge states were calculated with both semi-local (PBE) and hybrid DFT (HSE06), with poor agreement for non-fully-ionised charge states, but excellent agreement ($\Delta = \SI{0.02}{eV}$) for split-vacancy \kvc{V}{Al}{-3}.
Split vacancies have mostly only been reported for defects in their fully-ionised charge states (\cref{table:split_vacancy_energies}), and so we see that semi-local DFT is in most cases sufficient to describe the energetic preference for split over simple vacancies in these cases.\\

\begin{figure}[h]
\centering
\includegraphics[width=\textwidth]{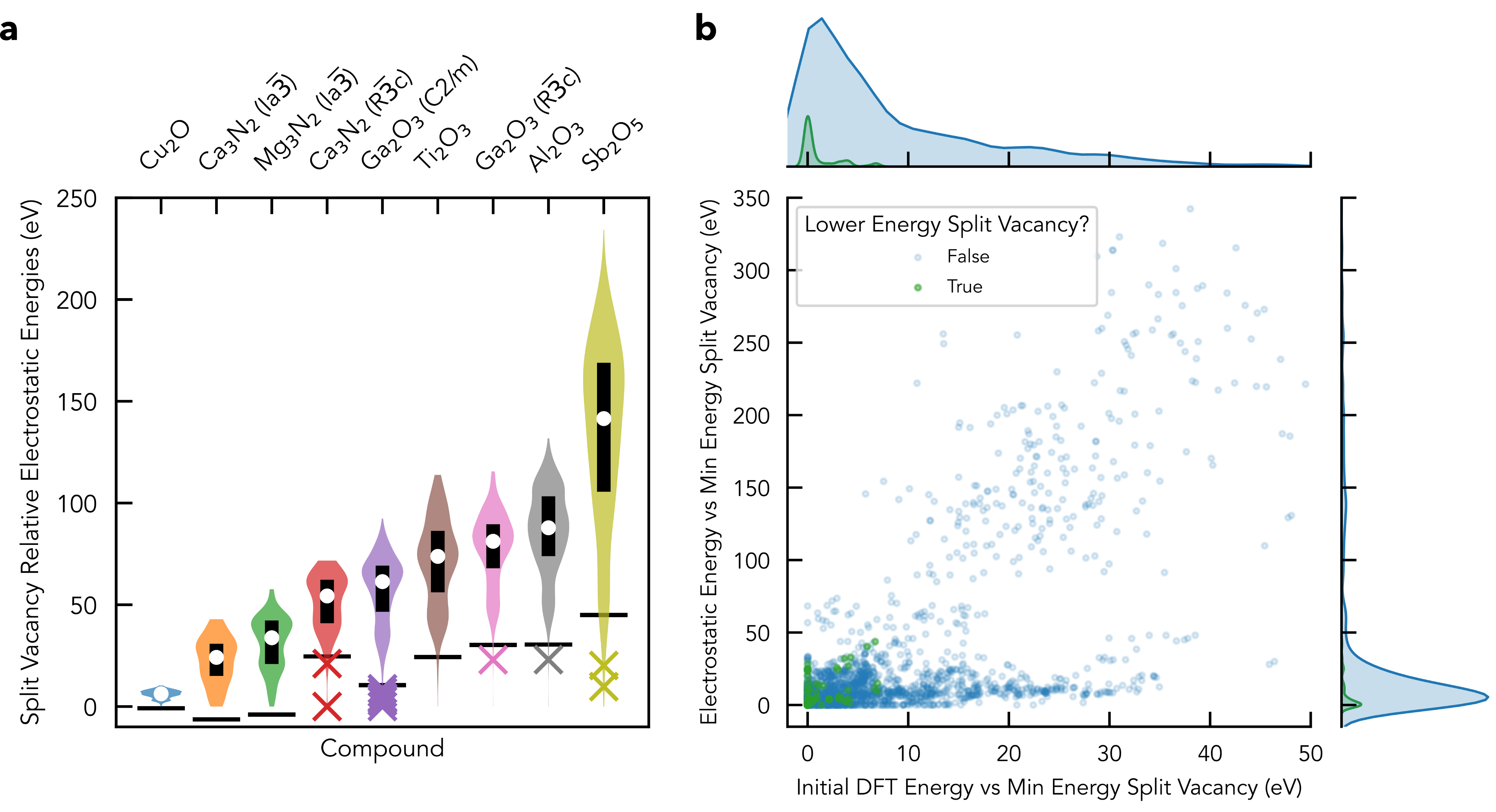}
\caption{
% \textbf{If we end up wanting to put the structure example images in here, then this could be good as a 4-part square figure, more space etc}\\
Electrostatic and DFT energy distributions of investigated vacancy structures. 
\textbf{(a)} Violin distribution plots of the relative electrostatic energies of candidate \kv{V}{X}-\kv{X}{i}-\kv{V}{X} complexes, with \kv{V}{X}-\kv{X}{i} distances less than \SI{5}{\angstrom} and interstitial sites determined by Voronoi tessellation,\cite{kavanagh_doped_2024,kononov_identifying_2023} across the same initial compound test set as \cref{fig:Geometric_and_PBE0}b (\kv{V}{Cation} for oxides and \kv{V}{Anion} for nitrides). There are $\sim$500 candidate configurations in each set. White circles and black rectangles denote the median and inter-quartile range respectively. Short horizontal black lines indicate the chosen cut-off energy for further screening, corresponding to the \SI{110}{\%} of the simple electrostatic energy difference of the simple vacancy and pristine bulk supercells.% electrostatic formation energy of the lowest energy simple vacancy.
As in \cref{fig:Geometric_and_PBE0}b, split vacancies which are lower energy than the lowest energy simple vacancy are denoted by $\times$ symbols. 
Note that some compounds (\ce{Cu2O}, Ia$\bar{3}$-\ce{Ca3N2}, Ia$\bar{3}$-\ce{Mg3N2}, \ce{Ti2O3}) do not have lower energy split vacancies.
% Compounds which do not show lower energy split vacancy geometries are indicated with black stars.
\textbf{(b)} Joint distribution plot of the (initial) electrostatic energies of all candidate split vacancies in the full DFT calculated dataset ($\sim$1000 compounds) against their corresponding (initial) DFT energies, relative to the minimum energy candidate in both cases. These energies are computed for candidate split vacancies before performing geometry relaxation (as required for electrostatic screening).
Configurations which relax to split vacancies which are lower energy than the lowest energy point vacancy are highlighted in green.
}
\label{fig:electrostatic_energies_analysis}
\end{figure}

\begin{figure}[h]
\centering
\includegraphics[width=\textwidth]{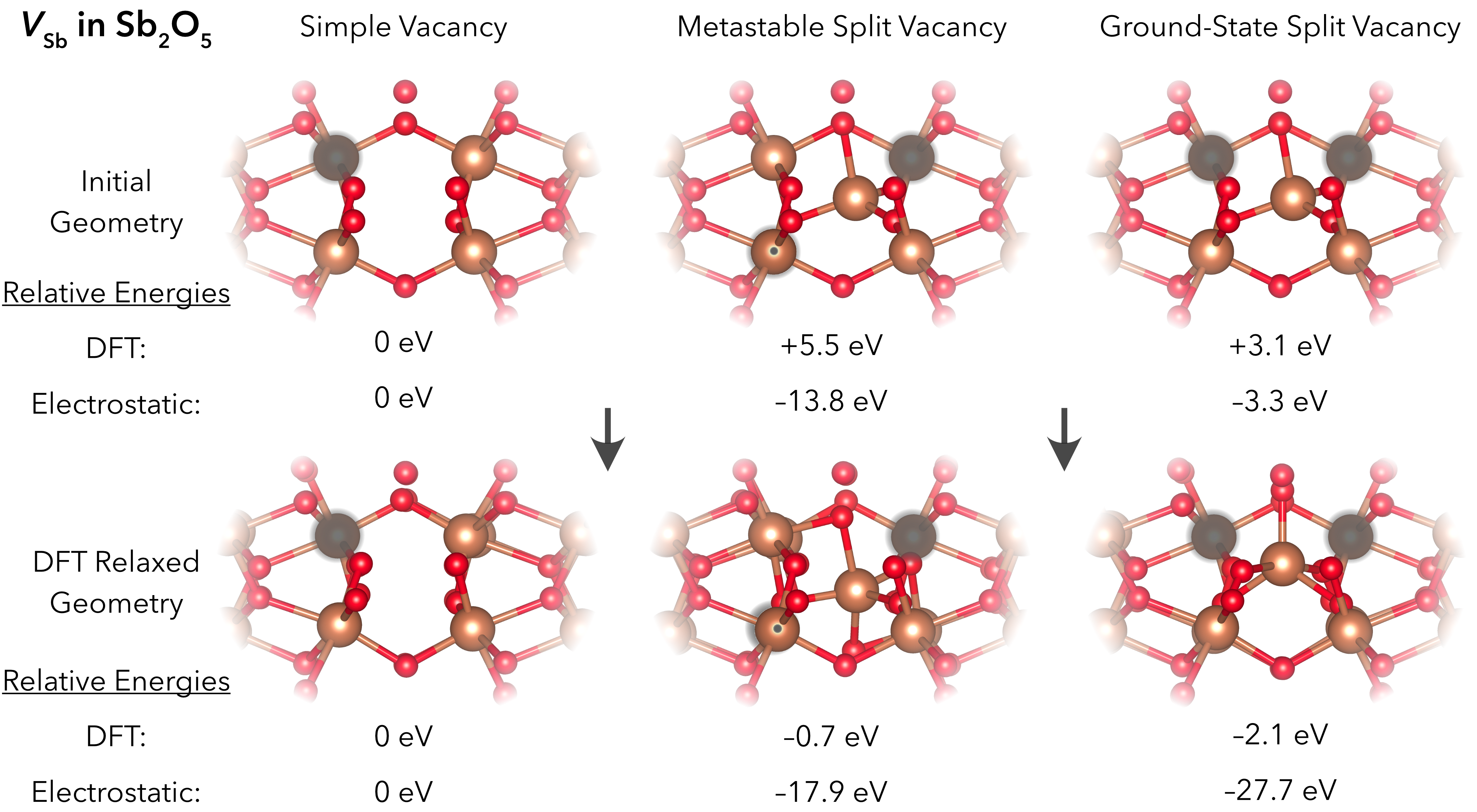}
\caption{Low energy vacancy configurations in \ce{Sb2O5}, before and after DFT relaxation. The relative energies according to DFT (PBEsol) and an electrostatic model (assuming formal ionic charges, inflating magnitudes) are shown alongside, with the simple point vacancy set to \SI{0}{eV} in each case. Vacancy positions are indicated by the semi-transparent grey circles.
}
\label{fig:Sb2O5_Example}
\end{figure}

Combined, the above considerations indicate that electrostatic and strain effects are the key driving factors for split vacancy formation in solids, with charge localisation and covalent bonding having minimal impacts.
I note that split vacancy geometries could certainly be favoured due to non-electrostatic / covalent bonding effects in some cases, but these are mostly unknown and likely rare.
% This suggests that consideration of cation-anion bonding in simple vs split vacancy configurations could be used to estimate the preference for split vacancy formation, prior to full quantum-mechanical calculation.  % Cut for brevity
With this in mind, I trial a simple electrostatic model using an Ewald Summation to compute Madelung energies, assuming all ions are in their formal charge states (i.e. fully-ionised charge states) and adding a compensating background charge density to avoid divergence, as implemented in the \texttt{pymatgen} \texttt{EwaldSummation}\cite{toukmaji_ewald_1996,ong_python_2013} tool.
\cref{fig:electrostatic_energies_analysis}a shows the distribution of these electrostatic energies for \kv{V}{X}-\kv{X}{i}-\kv{V}{X} complexes (example in \cref{fig:Sb2O5_Example}) for a selection of compounds which have previously been investigated for split vacancy formation.\cite{fowler_metastable_2024,li_computational_2024,kononov_identifying_2023,frodason_migration_2023}
I note that the electrostatic energy range here is quite large due to the use of formal oxidation states for the ionic charges (i.e. neglecting screening, which would require prior DFT / electronic structure calculations), however this should not affect the qualitative trends.
We also witness wider energy ranges for more highly-charged systems (e.g. Sb$^{+5}$ in \ce{Sb2O5} vs Cu$^{+1}$ in \ce{Cu2O}) as expected.
This approach yields a wide set of candidate geometries, the vast majority of which have highly-unfavourable electrostatic energies.
However, we see that in each compound with split vacancies \emph{lower} in energy than any simple point vacancy, the corresponding initial \kv{V}{X}-\kv{X}{i}-\kv{V}{X} geometries (denoted by $\times$ symbols) have electrostatic energies in the minimum tail of these broad distributions, indicating that this cheap electrostatic calculation can be used to effectively screen for low energy split vacancies.
If we take, as a screening cut-off value, the simple electrostatic energy difference of the simple vacancy and pristine bulk supercells, plus \SI{10}{\%} to account for energy shifts due to strain and relaxation effects, we find that known lower energy split vacancies are mostly captured by this range (indicated by the black horizontal lines in \cref{fig:electrostatic_energies_analysis}a).
This vacancy-dependent cut-off gives between 2 and 8 candidate split vacancy geometries in each case, with an average of $\sim$4, corresponding to a tiny subset of the $\sim$500 possible \kv{V}{X}-\kv{X}{i}-\kv{V}{X} configurations each.\\

This correlation between the electrostatic and DFT energies is further demonstrated in \cref{sifig:Initial_Test_Set_SI_ES_Analysis} for the initial test set and in \cref{fig:electrostatic_energies_analysis}b for all semi-local DFT calculations performed in this study.
We see that low electrostatic energies mostly correspond to low DFT energies (though with the neglect of strain, pair repulsion and covalent bonding still yielding significant spread), and initial geometries yielding lower energy split vacancies corresponding to those with low electrostatic energies (and low initial DFT energies).
% by comparing the relative electrostatic energies with the relative DFT energies (prior to geometry relaxation),
While the initial DFT energies of these configurations, prior to relaxation, are themselves not a perfect indicator of final \emph{relaxed} relative energies, we do see that they provide a decent estimate of relative stabilities, as indicated by the joint distribution plot of initial and final (pre- and post-relaxation) energies in \cref{sifig:Initial_vs_Final_DFT_energies}, and fact that all identified lower energy split vacancies in \cref{fig:electrostatic_energies_analysis}b have low initial DFT energies.
This is exemplified by the case of \kvc{V}{Sb}{-5} in \ce{Sb2O5} in \cref{fig:Sb2O5_Example}, which shows the initial and final (relaxed) DFT and electrostatic energies of the simple point vacancy and 2 of the 10 candidate \kv{V}{X}-\kv{X}{i}-\kv{V}{X} complexes predicted.
Here the initial electrostatic energies do not perfectly correlate with the DFT energies, but are effective in reducing the search space by identifying low \emph{electrostatic energy} configurations which may yield low \emph{total energies} upon relaxation.
While not perfectly precise, we see that this simple geometric and electrostatic approach allows us to rapidly screen through hundreds of possible \kv{V}{X}-\kv{X}{i}-\kv{V}{X} configurations for each candidate vacancy and reduce this to a small handful of candidate geometries which are then tractable for DFT energy evaluation -- particularly given the demonstrated accuracy of cheaper semi-local DFT \emph{for these specific fully-ionised defects}. \\
% I note that the initial cation-oxygen bond lengths for the interstitial cation in the split-vacancy structure, before relaxation, are closer to the longer cation-oxygen bond lengths (\SI{2.07}{\angstrom}), and so prior to relaxation this
% electrostatically disfavoured but still initial DFT energies favour it
% however avoiding the full rupturing of the shorter cation-anion bonds is still found to be the energetically...
% Relaxation effects
% Glosses over the fact that the created bond lengths at the split site are actually originally \SI{2.07}{\angstrom} (so not as short)(but relaxes to 2 A...) but seems the other broken bonds can then better relax away when it's still 2 short 1 long rather than 2 long 1 short 
% Of the 6 remaining \emph{under-coordinated} oxygen anions, their cation coordination has now increased, with the distance to the 4 closest cations (4 valent O here) now decreasing (from 3 x \SI{2.33}{\angstrom} \& 3x \SI{2.36}{\angstrom} 
% distance to the nearest (non-bonded) cation has now decreased \SI{3.69}{\angstrom}

\subsection*{Screening Split Cation Vacancies in Oxides}
\begin{figure}[h]
\centering
\includegraphics[width=\textwidth]{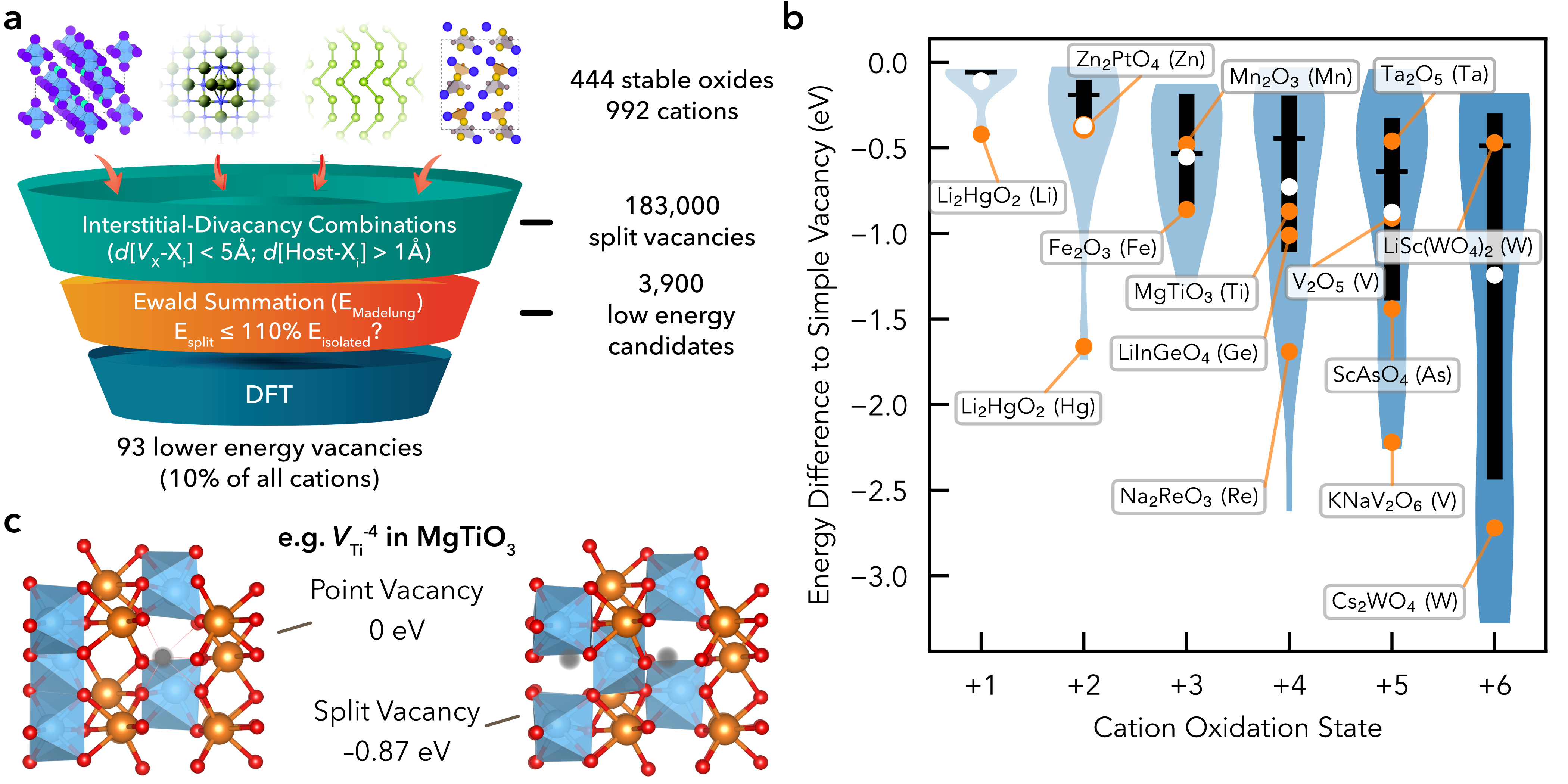}
\caption{Screening split cation vacancies in metal oxides. \textbf{(a)} Schematic diagram of the initial screening workflow employed to identify split cation vacancies in stable metal oxide compounds. 93 lower energy cation vacancies are identified, corresponding to $\sim$\SI{10}{\%} of all possible cation vacancies, and $\sim$\SI{20}{\%} of cation vacancies with at least one low electrostatic energy \kv{V}{X}-\kv{X}{i}-\kv{V}{X} arrangement. 
\textbf{(b)} Distribution of split vacancy energies relative to the lowest energy symmetry-inequivalent point vacancy for different cation oxidation states, for all lower energy split cation vacancies in metal oxides identified in this work ($\Delta E < \SI{-0.025}{eV}$). Some example compounds and the corresponding cation (vacancy) are shown as labelled orange datapoints. White circles, black dashes and rectangles denote the mean, median and inter-quartile range respectively.
\textbf{(c)} Example of DFT-relaxed point vacancy and split vacancy structures for \kvc{V}{Ti}{-4} in \ce{MgTiO3}, with Ti in blue, Mg in orange, O in red and vacancies as semi-transparent grey circles. Only Ti polyhedra are shown for clarity.
}
\label{fig:Kumagai_Oxides_Screening}  %  An example TLD could be nice...
\end{figure}

With these insights, I construct a workflow to screen for split vacancy formation, and apply it to a set of metal oxide compounds in order to test its efficacy and probe the prevalence of split vacancies in this important chemical space, as illustrated in \cref{fig:Kumagai_Oxides_Screening}a.
% With this efficient computational model for estimating split vacancy formation, I apply a high-throughput screening approach to identify split cation vacancies in metal oxides, 
Metal oxides are used in a wide variety of materials applications, such as transparent conducting oxides (TCOs), power electronics, battery cathodes, heterogeneous catalysis and more, for which defects play crucial roles.
Cation vacancies are typically the dominant acceptor defects in oxides,\cite{zhang_optimizing_2024,yuan_first-principles_2024}\footnote{~Outside of mixed-cation oxides with chemically-similar heterovalent cations.\cite{murdock_li-site_2024,hoang_defect_2014}} thus playing key roles in electronic conductivity --- counter-balancing the effects of positively-charged oxygen (anion) vacancies, ion diffusion, catalytic activity, optical properties and more.
% To this end, I apply this screening workflow to the database of $\sim$1000 stable insulating metal oxide compounds previously investigated by \citeauthor{kumagai_insights_2021}\cite{kumagai_insights_2021} for their oxygen vacancy properties.
Here I focus on the set of stable insulating metal oxides previously investigated by \citeauthor{kumagai_insights_2021}\cite{kumagai_insights_2021} for their oxygen vacancy properties.
Briefly, this dataset comprises metal oxide compounds on the Materials Project\cite{jain_materials_2013}\cite{kumagai_insights_2021} which are predicted to be thermodynamically stable against competing phases, are dynamically stable with respect to $\Gamma$-point phonon modes (i.e. unit-cell-preserving distortions), have 4 or less symmetry-inequivalent oxygen sites, 30 or less atoms in their primitive cells (to avoid complex low symmetry structures), do not have multiple anions and do not have partially-occupied $d$ or $f$ orbitals.
Further details on this dataset are available in Ref. \citenum{kumagai_insights_2021}.\\

\begin{algorithm}[h]
\small
\caption{Split Vacancy Screening Workflow. Steps in dashed boxes only apply to the screening of the full Materials Project\cite{jain_materials_2013} database.\\}
\label{alg}
\textbf{Input:} Host crystal structure \\
% \eIf{Host is charged (electrostatic driving force for split vacancies)}
\textbf{Supercell Generation:}
\begin{itemize}
    \item \underline{Metal Oxides:} Diagonal expansion of conventional unit cell to give most isotropic supercell with 60 $\leq$ Number of Atoms $\leq$ 500 --- taken from \citeauthor{kumagai_insights_2021}\cite{kumagai_insights_2021}
\end{itemize}
\begin{tikzpicture}
\node[rectangle,minimum width=\textwidth] (m) {\begin{minipage}{\textwidth}
\begin{itemize}
    \item \underline{Materials Project:} Scan all supercell expansions of primitive unit cell (including non-diagonal matrices), selecting the smallest supercell with minimum image distance $\geq$ \SI{10}{\angstrom} and $\geq$ 50 atoms --- \texttt{doped}\cite{kavanagh_doped_2024} algorithm
\end{itemize}
\end{minipage}};
\draw[dashed] (m.south west) rectangle (m.north east);
\end{tikzpicture}

\textbf{Candidate Generation:}
Generate all symmetry-inequivalent split vacancy geometries (\kv{V}{X} -- \kv{X}{i} -- \kv{V}{X} combinations) using \texttt{doped}\cite{kavanagh_doped_2024} under the constraints:
\begin{itemize}
    \item Both \kv{V}{X} -- \kv{X}{i} distances are $<$\SI{5}{\angstrom}
    \item Minimum distance from \kv{X}{i} to host lattice is $>$\SI{1}{\angstrom}
\end{itemize}
\textbf{Electrostatic Screening:}
Evaluate the electrostatic energy differences of all point \& split vacancies with the pristine bulk supercell ($\Delta E_{\textrm{electrostatic}}$), retaining only those with:
\begin{itemize}
    \item $\Delta E_{\textrm{electrostatic, split}}$ $\leq$ $\Delta E_{\textrm{electrostatic, point}} \times 1.10$ \\ (i.e. within \SI{110}{\%} of the electrostatic energy difference of the simple point vacancy and pristine bulk supercell)
\end{itemize}

\begin{tikzpicture}
\node[rectangle,minimum width=\textwidth] (m) {\begin{minipage}{\textwidth}
\textbf{Machine-Learned Interatomic Potential (MLIP) Screening:}
Evaluate the relative energies $E_{\textrm{MLIP}}$ of remaining split vacancy candidates using a machine-learned interatomic potential, retaining only those where:
\begin{itemize}
    \item $E_{\textrm{MLIP, split}}$ $\leq$ $E_{\textrm{MLIP, point}}$ (`standard' approach)
\end{itemize}
\textbf{\textit{or}}
\begin{itemize}
    \item $E_{\textrm{MLIP, split}}$ $\leq$ $E_{\textrm{MLIP, point}} + \SI{0.35}{eV}$ \textbf{\textit{and}} ML-relaxed structure retains split vacancy geometry (`exhaustive' approach)
\end{itemize}
\end{minipage}};
\draw[dashed] (m.south west) rectangle (m.north east);
\end{tikzpicture}

\textbf{DFT Evaluation:}
Compute the energies of the predicted split vacancy geometries using Density Functional Theory (DFT), to determine their stability relative to simple point vacancies in the given host structure.
\end{algorithm}

The screening workflow employed for this metal oxides dataset is summarised in \cref{alg} (without machine-learned potentials).
I first generate all symmetry-inequivalent \kv{V}{X}-\kv{X}{i}-\kv{V}{X} ($X$ = cation) complexes with \kv{V}{X}-\kv{X}{i} distances less than \SI{5}{\angstrom} using \texttt{doped}, prune to only those with electrostatic energy differences to the pristine bulk supercell $\leq$ \SI{110}{\%} of that for the lowest (electrostatic) energy simple vacancy (with a maximum of 10 per cation vacancy), and then calculate this subset with the PBEsol semi-local DFT functional, matching the calculation parameters used in the original database generation.\cite{kumagai_insights_2021}
Taking the first 444 metal oxide compounds in this database (comprising 992 possible cation vacancies), sorted by supercell size to optimise computational efficiency, this gives 183,000 possible \kv{V}{X}-\kv{X}{i}-\kv{V}{X} geometries, which is reduced to 3,900 using the electrostatic screening criterion (595 cation vacancies in 396 compounds).
This approach reveals 93 lower energy cation vacancies, corresponding to $\sim$\SI{10}{\%} of possible cation vacancies in this dataset, with energy differences to the lowest energy simple point vacancy between 0.05 and \SI{3}{eV} (\cref{fig:Kumagai_Oxides_Screening}b), with a mean energy lowering of \SI{0.81}{eV}.
Using a distance-based classification algorithm implemented in \texttt{doped}\cite{kavanagh_doped_2024} (\cref{SI:Split_Vacancy_Classification}), just over \SI{50}{\%} of the lower energy vacancies are determined to form split-vacancy type geometries like that shown in \cref{fig:Kumagai_Oxides_Screening}c, while the others rearrange to adopt distorted point vacancy structures which are lower energy than those obtained from standard geometry relaxations of the simple point vacancies.
These are significant energy differences, with the mean energy lowering $\overline{\Delta E} = \SI{0.81}{eV}$ amounting to over 3 orders of magnitude difference in equilibrium defect concentrations for a growth temperature of $T \sim$ \SI{1000}{K}, or 10 orders of magnitude in equilibrium populations at room temperature (relevant for charge compensation). \\

Moreover, this screening identifies many low-energy metastable vacancy geometries (e.g. as for \ce{Sb2O5}, \cref{fig:Sb2O5_Example}), finding 210 distinct metastable states with energies within \SI{0.5}{eV} of the lowest energy simple point vacancy, in 160 of the 600 cation vacancies which gave candidate low-energy sites from electrostatic screening, with distributions and tabulated data provided in \cref{SI:Meta_States}.
Here I classify distinct metastable states as those which (i) relax to a split vacancy geometry (determined by the \texttt{doped}\cite{kavanagh_doped_2024} classification algorithm), with no corresponding symmetry-inequivalent point vacancy spontaneously relaxing to a split vacancy, and (ii) are different in energy by $>$\SI{25}{meV} to all other metastable states for that vacancy.
Identification of low-energy metastable states for defects in solids is important for understanding a number of key defect properties, such as carrier recombination rates,\cite{kavanagh_rapid_2021,kavanagh_impact_2022,alkauskas_role_2016} oxidation and decomposition,\cite{squires_oxygen_2024,cen_cation_2023} catalytic activity,\cite{tan_unravelling_2020,zhang_toward_2023} field-effect transistors\cite{grasser_two-stage_2009} and more.\cite{lee_investigation_2024,frodason_migration_2023}
Many of these compounds which exhibit lower energy split vacancies are being investigated for functional materials applications where defect behaviour is crucial, such as \ce{MgTiO3} which is used in wireless communication for its excellent dielectric properties (\cref{fig:Kumagai_Oxides_Screening}c),\cite{kuganathan_defects_2019,magar_advancing_2024} \ce{CsReO4} for potential applications in photocatalysis and radioactive waste storage,\cite{mullens_tetrahedra_2024} \ce{Ta2O5} for oxygen evolution reaction (OER) catalysis,\cite{sathasivam_computational_2017,han_understanding_2024} \ce{Fe2O3} for photoelectrochemical water oxidation,\cite{zhao_-fe2o3_2021,banerjee_thermokinetics_2023} and \ce{V2O5} for high capacity battery electrodes.\cite{boruah_light_2021,sucharitakul_v2o5_2017}  % \ce{Li⁢In⁢Ge⁢O4} for laser action,\cite{sharonov_spectroscopy_2005} % \ce{Li2GeO3} for unusual optical activity,\cite{dien_orbital-hybridization-created_2021}
For instance, the preference for Co/Ni migration to interlayer positions in delithiated \ce{Li$_{1–x}$(Co,Ni)$_x$O2} cathode materials, known to drive degradation and capacity fade,\cite{squires_oxygen_2024,vinckeviciute_fundamental_2019,genreith-schriever_probing_2024} is predicted here -- corresponding to split Co/Ni vacancies in fully-delithiated \ce{Li$_{1–x}$(Co,Ni)$_x$O2}.
It is imperative that the correct ground and low-energy metastable states are identified in defect investigations of these compounds, as these large energy differences and distinct geometries could drastically affect predictions.\\

From these results, we can conclude that split vacancy defects do have a significant prevalence across different materials and structure types, and are not just limited to the handful of known cases discussed in the introduction --- which are mostly (anti-)corundum structures.
As expected, larger cation oxidation states give larger ranges of energy lowering (\cref{fig:Kumagai_Oxides_Screening}b), corresponding to stronger electrostatic bonding and thus greater energy variations with different ion arrangements (matching the trends in electrostatic energy ranges in \cref{fig:electrostatic_energies_analysis}a).
These large energy-lowering reconstructions for negatively-charged cation vacancies in oxides will make these acceptor defects much shallower, yielding greater compensation of any $n$-type doping -- typically from oxygen vacancies.\cite{kumagai_insights_2021}
These lower energy geometries are also expected to inhibit cation migration, having increased energy barriers to displacement away from the equilibrium geometries, as for \ce{Ga2O3}.\cite{frodason_migration_2023,varley_hydrogenated_2011}

\subsection*{Machine Learning Acceleration}
From the above results, we see that this geometric and electrostatic pre-screening model greatly reduces the search space for split vacancy configurations in solids, allowing the screening and identification of these species in several hundred metal oxides.
This once again highlights that the primary contributions to low energy split vacancy formation are electrostatic effects, with the remaining inaccuracies (prior to DFT computation) stemming from strain, pair repulsion and remnant covalent bonding effects which are not captured by this simple model.
These considerations hint at the possibility of using some form of simple energy potential to estimate these contributions to further improve the accuracy and thus efficiency of this pre-screening approach.
Machine-learned force fields (MLFFs) -- machine learning models trained on energies and forces from quantum-mechanical simulations -- present an attractive option for this goal.\cite{mosquera-lois_machine-learning_2024}
In particular, `foundation' MLFFs (a.k.a. universal potentials) are trained on large and diverse datasets of DFT simulations, affording generality to these models (applicable to compositions spanning the periodic table) and achieving accuracies close to that of semi-local DFT but at a small fraction of the computational cost.\cite{batatia_foundation_2024,riebesell_matbench_2024,deng_chgnet_2023}\\

Here, I take the \texttt{MACE-mp} foundation model,\cite{batatia_foundation_2024,batatia_mace_2022} which is an equivariant graph neural network force field trained on semi-local DFT (PBE) geometry relaxations for inorganic crystalline solids in the Materials Project database.\cite{jain_materials_2013}
Using this universal potential to relax all split vacancy candidate geometries in the metal oxides test set discussed above, I find that it successfully predicts the energetic preference for split vs simple vacancies in \SI{88.1}{\%} of the $\sim$600 cation vacancies calculated with DFT.
Inaccuracies in the foundation model predictions are more concentrated in cases where split vacancies are the \emph{lowest} energy geometry however, with \texttt{MACE-mp} correctly predicting the split vacancy state for \SI{53}{\%} of the energy-lowering split vacancies subset.
If we take a more exhaustive approach, where we consider distinct \texttt{MACE-mp} relaxations yielding a split vacancy geometry and energy within \SI{0.35}{eV} of the ground state as candidates (to maximise our true positive hit rate, at the cost of some efficiency (i.e. false positives)), this gives a model which correctly identifies the split vacancy state for \SI{81}{\%} of our energy-lowering split vacancies subset.
The performance metrics of the best models are tabulated in \cref{table:MACE_performance}. 
Here we achieve an F1 score of 0.6 --- a commonly-used ML classification metric --- and a `discovery acceleration factor' (DAF)\cite{riebesell_matbench_2024} of $\sim120$ --- which represents the computational speedup in discovery provided by the ML model, with a maximum possible value of 1/Prevalence $\simeq$ 143 --- and true positive rates (TPR) up to \SI{81}{\%}.\\

\AtBeginEnvironment{table}{\mathindent=0pt}  % don't indent figure caption, default in IOP format
\begin{table}[h]
\def\arraystretch{1.25}  % cleaner, more space
\caption{Performance metrics for \texttt{MACE-mp} foundation models in identifying split vacancy configurations for the test set of cation vacancies in metal oxides (\cref{fig:Kumagai_Oxides_Screening}), with candidate geometries from geometric \& electrostatic screening as input. The prevalence rate of candidate geometries for which a split vacancy is the lowest energy state (with $\Delta E < \SI{-25}{meV}$) is \SI{0.7}{\%}.
% \SI{8}{\%} for prevalence rate of _cation vacancies_ which give lower energy split vacancies...  -- the latter indicated by `prevalence$^*$' here. 
F1 score is a common metric used for ML classification methods, while the `discovery acceleration factor' (DAF) quantifies the computational speedup in discovery (of split vacancies in this case) compared to random selection.\cite{riebesell_matbench_2024}
DAF has a maximum value of 1/Prevalence $\simeq$ 143, for perfect accuracy.  % 12.5 
For TPR/FPR/TNR/FNR; T = True, F = False; P = Positive, N = Negative; R = Rate. 
The definitions of the various metrics are given under their title.
`Small' refers to the \texttt{MACE-mp-small} model, while `exhaustive' is when ML-predicted metastable split vacancies with $\Delta E < \SI{0.35}{eV}$ are included (see text). Champion values shown in bold.
}
\begin{center}
\resizebox{\textwidth}{!}{%
\centering
\begin{tabular}{l|c|c|c|c|c|c|c|c}
\hline  % And underline which is best for each metric?
\textbf{Model} & 
\begin{tabular}[t]{@{}c@{}}\textbf{F1} \\ $\frac{TP}{TP + (FP + FN)/2}$ \end{tabular} & 
\begin{tabular}[t]{@{}c@{}}\textbf{Precision} \\ $\frac{TP}{TP + FP}$ \end{tabular} & 
% \begin{tabular}[t]{@{}c@{}}\textbf{DAF} \\ $\frac{\textrm{Precision}}{\textrm{Prevalence}}$ \end{tabular} & 
\begin{tabular}[t]{@{}c@{}}\textbf{DAF} \\ $\frac{\textrm{Precision}}{\textrm{Prevalence}^*}$ \end{tabular} & 
\begin{tabular}[t]{@{}c@{}}\textbf{TPR} \\ $\frac{TP}{TP + FN}$ \end{tabular} & 
\begin{tabular}[t]{@{}c@{}}\textbf{FPR} \\ $\frac{FP}{FP + TN}$ \end{tabular} & 
\begin{tabular}[t]{@{}c@{}}\textbf{TNR} \\ $\frac{TN}{TN + FP}$ \end{tabular} & 
\begin{tabular}[t]{@{}c@{}}\textbf{FNR} \\ $\frac{FN}{FN + TP}$ \end{tabular} \\
\hline
small & \textbf{0.63} & \textbf{0.78} & \textbf{118.9} & 0.53 & \textbf{0.01} & \textbf{0.99} & 0.47 \\  % \textbf{9.76} &
% \textbf{ensemble} & 0.63 & 0.78 & 9.76 & 0.53 & 0.01 & 0.99 & 0.47 \\
exhaustive & 0.56 & 0.43 & 64.9 & \textbf{0.81} & 0.09 & 0.91 & \textbf{0.19} \\  % 5.33 & 
\hline
\end{tabular}
}
\end{center}
\label{table:MACE_performance}
\end{table}  % Ensemble model gives 93.2\% (where they agree in at least one case)

These values show that the ML foundation model can provide a 2 orders of magnitude speedup in identifying low energy split vacancies, with reasonable accuracies.
% The high DAF values here demonstrate the significant speedup offered by the ML foundation model in identifying low energy split vacancies.
% Given that I only take the lowest energy split vacancy predicted by the ML model (rather than all candidate geometries from the electrostatic screening step, with an average of 12.2 per cation vacancy), the actual acceleration factor in terms of numbers of DFT calculations to perform is DAF$^*$ = 118.9, giving over two orders of magnitude speedup.
Moreover, these discovery metrics are based on the prevalence of lower energy split vacancies within the electrostatically-screened geometries, and so the actual acceleration factor of this electrostatic \& ML approach compared to random selection of all possible \kv{V}{X}-\kv{X}{i}-\kv{V}{X} geometries would be orders of magnitude larger.
Applying the `small' model to the remaining fraction of the metal oxides dataset\cite{kumagai_insights_2021} and then using DFT to compute the energies of predicted lower energy split vacancies, relative to the symmetry-inequivalent point vacancies, I find that it successfully predicts a \emph{lower} energy split vacancy in \SI{44}{\%} of cases.
Notably, the majority of `false positives' here do retain split vacancy geometries, adopting low energy metastable states which are still relevant to defect investigations as discussed above.\cite{lee_investigation_2024,frodason_migration_2023,fowler_metastable_2024}
I note that a number of model choices, such as model size, floating point precision, geometry optimisation algorithm and more, were tested for this step in the workflow, as detailed in \cref{SI:MACE_tests}.\\
% Note the hit rate of our DFT calculations? with simple geometric model, simple ES, ES + MACE? Idk, not so easy to define
% It should be noted that the \texttt{MACE-mp} model was trained on PBE DFT calculations, while the metal oxides test set here was calculated with PBEsol, and so some model inaccuracies here could possibly   % Nah tbf for our split vacancy failure cases, doesn't seem like small PBE vs PBEsol differences would really explain it
% \textbf{Note that some differences/failures/discrepancies with MACE and our DFT results could also be due to comparing PBEsol and a PBE trained model... (also we didn't set MAGMOM for the tungsten)}

\begin{figure}[h]
\centering
\includegraphics[width=\textwidth]{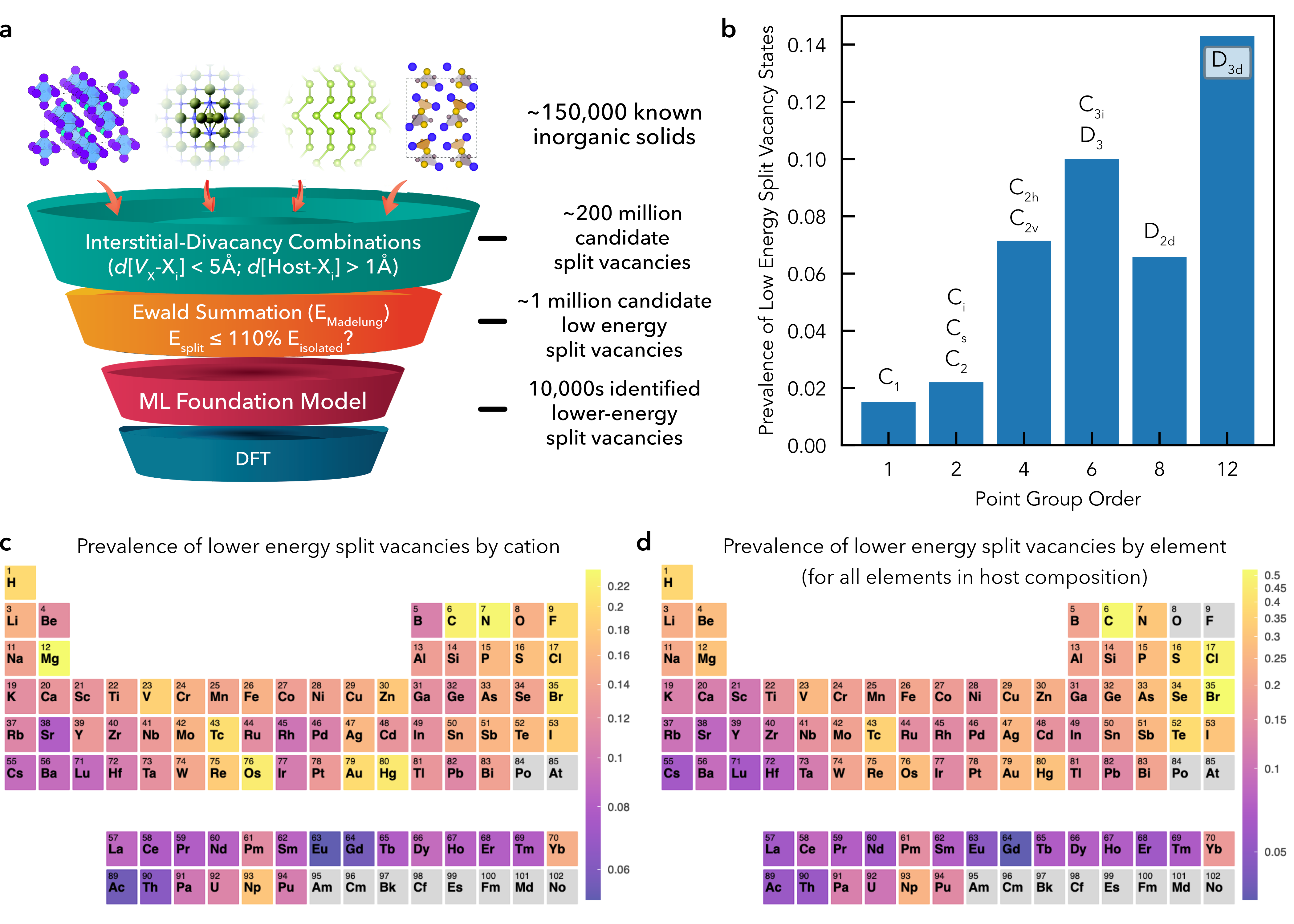}
\caption{
% Space groups not so clear, point group is however
ML-accelerated screening of split vacancies. 
\textbf{(a)} ML-accelerated screening workflow employed to predict the formation of split cation vacancies in all compounds in the Materials Project\cite{jain_materials_2013} (MP) database. 
\textbf{(b)} Normalised prevalence of distinct low-energy split vacancies within the electrostatically-screened set of candidate \kv{V}{X}-\kv{X}{i}-\kv{V}{X} geometries for all cation vacancies in metal oxides computed with DFT, as a function of the point group order of the \kv{X}{i} site (as given by the \texttt{doped}\cite{kavanagh_doped_2024} defect symmetry functions). 
Values are normalised by the total prevalence of each point group within candidate \kv{X}{i}. %  the electrostatically-screened set of 
The point group order corresponds to the number of symmetry operations, with higher order corresponding to higher symmetry.
\textbf{(c,d)} Prevalence of lower energy split vacancies, as predicted by the ML-accelerated screening of the MP database, for \textbf{(c)} the cation vacancy element and \textbf{(d)} all elements in the corresponding host compound.
Values are normalised by the elemental prevalences in the dataset, and a logarithmic colourbar scale is used. The same heatmaps, weighted by energy lowering magnitudes, are provided in \cref{sifig:Weighted_Prevalence_Plots}. 
Generated using the \texttt{periodic\_trends}\cite{noauthor_andrew-s-rosenperiodic_trends_nodate} script.
}
\label{fig:MP_Screening}
\end{figure}

This predictive ability of the foundation ML model, on top of the geometric and electrostatic screening step, allows an accelerated tiered screening procedure (summarised in \cref{alg}, including the ML step) to estimate the formation of low energy split vacancies, making it applicable to extremely large datasets of materials.
As depicted in \cref{fig:MP_Screening}, I use this approach to search for split cation vacancies in all compounds on the Materials Project (MP) database, which includes all entries in the ICSD (as of November 2023), along with several thousand computationally-predicted metastable materials.\cite{jain_materials_2013}
With this screening procedure, the ML model predicts lower energy structures missed by standard defect relaxations for $\sim$55,000 (\SI{20}{\%}) cation vacancies in 43,000 (\SI{29}{\%}) materials, of which 29,000 (\SI{10}{\%}) are classified as split vacancies.\\
% --- where the percentages in parentheses are given by normalising by the total number of cation vacancies / materials in the dataset.\cite{kavanagh_doped_2024}

To validate the accuracy of this approach, and investigate the prevalence of split cation vacancies within another chemical subspace, I take all cases of ML-predicted lower energy split vacancies in nitride compounds which are thermodynamically-stable (according to MP calculations) and do not contain lanthanides (which can be poorly modelled by standard DFT), and again compute the energies of predicted lower energy split vacancies relative to the symmetry-inequivalent point vacancies using DFT, matching the MP computational setup.
Here the ML model predicts 198 (103) cation vacancies with lower energy structures, while DFT shows this to be true in 113 (40) cases, corresponding to a predictive accuracy of \SI{57}{\%} (\SI{39}{\%}) here -- values in parentheses corresponding to those where relaxed geometries are classified as split vacancies.
The slightly worse predictive accuracy for nitrides (\SI{39}{\%}) compared to oxides (\SI{44}{\%}) is likely a result of the greater prevalence -- and thus expected accuracy for -- oxides in the MPTrj dataset upon which \texttt{MACE-mp} is trained.\cite{riebesell_matbench_2024}
If we take the lower-bound predictive accuracy of $\sim$\SI{40}{\%} from our oxide and nitride test sets, this would correspond to $\sim$22,000 (12,000) true lower energy cation vacancy structures being identified by the electrostatic + ML screening of the MP database, and around \SI{60}{\%} more if we employ the `exhaustive' model.\\
% Including the \SI{0.35}{eV} uncertainty region, which previously showed a \textbf{X} hit rate, gives \textbf{Y} potentially-low-energy split cation vacancies...
% To further validate the effectiveness of this approach, I take all nitride compounds which are predicted to be thermodynamically stable (having zero energy above hull according to the Materials Project\cite{jain_materials_2013} database), and compute the energies of their simple and low energy split vacancy configurations predicted by the MACE ML foundation model in this screening.
% This corresponds to 1311 compounds, for which \textbf{Z} are predicted to exhibit low energy split cation vacancies by this screening method.
% Using the PBE semi-local DFT functional to match the Materials Project calculation setup,\cite{jain_materials_2013} I find that \textbf{U} of these cases do yield...
% \textbf{Some quick analysis of failure cases?}
% \textbf{If time, would be nice to pick a specific example and show the effect on the full TL diagram, at the hybrid level, demonstrating the importance and massive effect this can have. Look at our known set and decide for this – pick one from the Kumagai db to illustrate, prob best? But maybe doesn't matter because we don't have hybrid structures... use one of the nitrides?}

% \textbf{Some side by side distribution plots could be good (oxides and nitrides, like softening paper, or side-by-side bar charts), then also need to do some for the MP scan as well? Periodic table heatmap would be good too}
Analysing the distributions of DFT-confirmed low energy split cation vacancies in metal oxides, no clear trends are seen for the host compound space groups, however we do see a trend when looking at point symmetries.
\cref{fig:MP_Screening}b shows that higher symmetry geometries, as measured by the point group order of the \kv{X}{i} site in the \kv{V}{X}-\kv{X}{i}-\kv{V}{X} geometry, correlate with a higher prevalence of low energy split vacancies.
This is likely due to higher symmetry sites corresponding to greater cation-anion bonding and more bulk-like coordination, favouring lower energies as seen in the example cases discussed in the introduction.
\cref{fig:MP_Screening}c shows that certain cations have particularly high prevalences of predicted lower energy split vacancies, such as cationic carbon, chalcogens, halogens, hydrogen, magnesium, mercury, vanadium and coinage metals (Cu, Ag, Au).
Looking at host compound compositions (\cref{fig:MP_Screening}d), we see that compounds with nitrogen, carbon, magnesium, bromine, osmium and mercury have high prevalences of predicted lower energy split vacancies, while prevalence rates decrease on average as we move down the periodic table.
% \textbf{Clearly prevalent, but mostly in certain compositions subsets. Depends a lot of material, oxi state, symmetry, stoichiometry etc?}
% \textbf{If time: Should analyse some outliers, seems like maybe certain cations are trickier to guess than others for MACE, could indicate poor training / sparsity in the MPTrj training set. Tl seems quite poorly modelled for instance? (Was likely the +/-1 issue)}

\section*{Discussion \& Conclusions}
The identification of ground and metastable configurations is crucial to our understanding of defects in materials.
The defect geometry is the foundation from which its behaviour and associated properties (such as formation energy, concentration, migration, doping etc) derive, and is thus key to both experimental and theoretical characterization of defects.
For instance, some key impacts of split cation vacancy formation are the increased acceptor defect concentration (due to the significant lowering of the cation vacancy (dominant acceptor) formation energies), yielding greater compensation of oxygen vacancy donors and impacting the migration of highly-charged defects under strong electric fields.\cite{varley_hydrogenated_2011,frodason_migration_2023}
The importance of energy-lowering reconstructions from simple, unperturbed defect geometries to key material properties has been demonstrated in a wide range of materials and applications in recent years, aided by improved structure-searching methods,\cite{arrigoni_evolutionary_2021,mosquera-lois_shakenbreak_2022,mosquera-lois_machine-learning_2024} however `non-local' defect reconstructions such as split vacancies remained elusive.
We see that these defect species are present in significantly more materials than previously known (around \SI{10}{\%} of cation vacancies in all inorganic solids), with prevalence and importance varying as a function of composition, point symmetries and oxidation states.
This once again highlights the importance of advanced global optimisation methods for defects which can accurately and efficiently scan their potential energy surfaces, to identify which defects are actually present.\cite{mosquera-lois_search_2021}\\

To rapidly assimilate the findings of this study to a directly usable form for the defect community, this database of screened split vacancy configurations in known inorganic solids has been made directly accessible through the \texttt{doped}\cite{kavanagh_doped_2024} defect simulation package. 
\texttt{doped} automatically queries this database upon defect generation, informing the user if low energy split vacancies are predicted for the input host compound, and at what level of confidence.
This will allow defect researchers to easily and automatically incorporate the behaviour of low energy and metastable split vacancies in future studies, boosting the accuracy of predictions. % without the need to perform advanced structure-searching, --- as a community
Moreover, all code used to generate and screen candidate split vacancies, using \texttt{doped},\cite{kavanagh_doped_2024} \texttt{pymatgen},\cite{ong_python_2013} \texttt{vise},\cite{kumagai_insights_2021} \texttt{ASE},\cite{larsen_atomic_2017} and \texttt{MACE}\cite{batatia_mace_2022,batatia_design_2025}, along with the full database of predicted structures is openly available at \href{https://doi.org/10.5281/zenodo.XXXX}{https://doi.org/10.5281/zenodo.XXXX} \textbf{[released upon publication]}. % and confidence of low energy split vacancy formation 
The individual functions to implement each step in the structure-searching and screening approach from this work are likewise implemented in \texttt{doped}\cite{kavanagh_doped_2024}, so that users can predict the formation of split vacancies in novel compounds not listed on the Materials Project\cite{jain_materials_2013} database, make use of future ML models with improved accuracy/efficiency for screening, or examine the possibility of other low-energy complex defects with a similar approach.\\\\

It is worth noting that these split vacancies are essentially a stoichiometry-conserving defect complex; \kv{V}{X}-\kv{X}{i}-\kv{V}{X}; rather than a true `point defect', which raises some general questions and considerations for their interpretation.
% In this way, we can equally think of their formation as a transformation from a single simple vacancy to the split vacancy cluster as in \cref{fig:Split_Vacancies_Intro_Fig}, or as a tightly-bound complex formation from the clustering of the constituent point defects (\kv{X}{i} \& \kv{V}{X}).
% Of course in reality this is just a quasiparticle-like\cite{de_souza_revisiting_2023} picture we use to aid in the interpretation and characterisation of defects, which often form during crystal growth from highly random initial atomic arrangements. % not sure if this is useful
% It does, however, raise some general questions and considerations for their interpretation.
For instance, their multiplicity and degeneracy pre-factors ($Ng$) in the defect concentration equation ($N_X = Ng\exp{(\sfrac{-\Delta E}{k_B T})}$) will in most cases be reduced from that of the simple point vacancies.
In $\alpha$-\ce{Al2O3} for example, there is only one split-vacancy configuration per 2 host Al sites (i.e. possible simple vacancy sites), reducing its multiplicity pre-factor by half.\cite{mosquera-lois_imperfections_2023,kavanagh_impact_2022}\footnote{~These degeneracy factors and complex defect multiplicities are automatically computed and incorporated in thermodynamic analyses by \texttt{doped}.\cite{kavanagh_doped_2024}}
Indeed, often the energy-lowering reconstructions we observe for point defects can be thought of as effectively `defect clusters' rather than true point defects.
% whether with \texttt{ShakeNBreak}\cite{mosquera-lois_shakenbreak_2022} or otherwise,
For instance, the famous original cases of `DX-centres' correspond to substitutional defects \kv{Y}{X} displacing significantly off-site, effectively transforming to interstitial-vacancy complexes \kv{Y}{i}-\kv{V}{X}.\cite{chadi_energetics_1989}
We have seen similar energy-lowering reconstructions to stoichiometry-conserving defect clusters in many cases with \texttt{ShakeNBreak},\cite{mosquera-lois_identifying_2023,mosquera-lois_shakenbreak_2022} such as the formation of dimers and trimers at vacancies chalcogenides\cite{wang_four-electron_2023,wang_sulfur_2025} and oxides\cite{li_computational_2024,squires_oxygen_2024} as neighbouring under-coordinated atoms displace toward the vacant site in essentially \kv{V}{X} $\rightarrow$ \kv{V}{Y} + \kv{Y}{X} transformations.
Advanced methods for identifying and characterising these defect clusters, similar to the vacancy classification algorithms used in \citeauthor{kumagai_insights_2021}\cite{kumagai_insights_2021} and in this work, will improve our understanding of the prevalence and typical behaviours of such defect clusters.
% , as well as helping to connect the two limiting cases of their interpretation; from single point defects or from multiple isolated constituent point defects.  % could prob be better worded, maybe don't need to say so much on this here
\\

The workflow introduced here could be effectively applied to screen for other low-energy defect complexes in solids.  % both stoichiometry-conserving and otherwise
Defect complexes typically involve fully-ionised constituent point defects and are mostly governed by electrostatic attraction and strain,\cite{krasikov_defect_2017} as was shown to be the case for split vacancies here. 
As such, they correspond to a similar computational problem of large configuration spaces but with relatively simple energetics, for which we see that geometric, electrostatic and universal MLFF screening can be powerfully applied.  % could cite Kanta paper here if preprinted
For instance, most single-photon emitters, with applications in quantum sensing, communication and computing, are complex defects such as the NV centre in diamond\cite{razinkovas_photoionization_2021} or the T centre in silicon.\cite{xiong_computationally_2024}
Indeed, efforts have already begun to computationally screen thousands of candidate complex defects to identify colour centres for quantum applications,\cite{xiong_computationally_2024,davidsson_na_2024} for which the screening approach and computational tools introduced here could be used to boost efficiency and scope.\\\\

This work serves as an exciting early demonstration of the power and utility of foundation ML potentials (a.k.a. universal MLFFs), allowing us to greatly expand the scope, scale and speed of computational materials investigations.
The non-locality and extremely large configuration space for these split vacancy defects presents a significant challenge for their identification.  %  (e.g. $\sim$500 candidate configurations per vacancy defect)
However, the fact that the underlying energetic driving factors are relatively simple (primarily electrostatics and strain), makes foundation MLFFs ideally suited to their identification. % despite this large configuration space, 
We see here that a general-purpose ML potential (MACE) is capable of predicting the formation of these species with reasonable accuracy.
% trained on compositions spanning most of the periodic table
Only a minute fraction of the candidate structures from electrostatic screening of the MP\cite{jain_materials_2013} dataset would have been possible to evaluate with DFT, whereas the foundation ML model can predict their relative energies with reasonable qualitative accuracy in the space of a day -- using a large number of GPUs.\\

Nevertheless, these exciting findings come with some important caveats.
These foundation models \emph{only} work so well here because the dominant bonding and energetics for these species are relatively simple, and do not involve carrier localisation, variable charges or excess charge (opposite to \emph{most} defect species and metastabilities/reconstructions),\cite{mosquera-lois_identifying_2023} for which they fail dramatically.
Their power in this case stems from the combination of an enormous configuration space with underlying energetics that are well-reproduced by the foundation model. % (trained on semi-local DFT calculations).
As such, this is a promising demonstration of the potential utility of machine learning (ML) approaches to defects and materials modelling in general, \emph{when used appropriately}, but as a field we are still a long way away from having generally-applicable ML methods for defects.\\

\section*{Acknowledgements}
I thank Irea Mosquera-Lois for a careful reading of this manuscript, valuable discussions regarding universal MLFFs, and for making useful parsing and plotting scripts\cite{mosquera-lois_machine-learning_2024} (and data) openly-available online. 
I acknowledge useful discussions with Dr Joel Varley regarding the performance of \texttt{ShakeNBreak} and other structure-searching strategies for split vacancy defects, Prof Beall Fowler regarding known split vacancy defects in solids, and Ke Li regarding split vacancies in \ce{Sb2O5} \& \ce{Al2O3}.
I would also like to acknowledge Prof Yu Kumagai for making his database of oxygen vacancy calculations in metal oxides openly-available online, which was used as an initial test set of compounds in this work.
I thank the Harvard University Center for the Environment (HUCE) for funding a fellowship.

\newcommand{\newblock}{}
\bibliographystyle{rsc} 
\bibliography{zotero}
% \putbib
% \end{bibunit}
%%%%%%%%%% Merge with supplemental materials %%%%%%%%%%
%%%%%%%%%% Prefix a "S" to all equations, figures, tables and reset the counter %%%%%%%%%%
\setcounter{equation}{0}
\setcounter{figure}{0}
\setcounter{table}{0}
\setcounter{section}{0}
\setcounter{secnumdepth}{3} % Section numbering back on
\setcounter{page}{1}
\makeatletter
\renewcommand{\theequation}{S\arabic{equation}}
\renewcommand{\thefigure}{S\arabic{figure}}
\renewcommand{\thetable}{S\arabic{table}}
\renewcommand{\thesection}{S\arabic{section}}
%\renewcommand{\bibnumfmt}[1]{[S#1]}
%\renewcommand{\citenumfont}[1]{S#1}
%%%%%%%%%% Prefix a "S" to all equations, figures, tables and reset the counter %%%%%%%%%%
% \begin{bibunit}  % section off these references for SI
%\newcommand{\green}[1]{{\leavevmode\color{black}{#1}}}  % for highlighting over multiple paragraphs

\clearpage
\title[Identifying Split Vacancy Defects with Foundation Models and Electrostatics (SI)]{Supplementary Material: Identifying Split Vacancy Defects with Machine-Learned Foundation Models and Electrostatics} % Try remember to update if any changes in title

\section{Additional Methodological Details}\label{SI:Additional_Methods}
\subsection{Split Vacancy Classification}\label{SI:Split_Vacancy_Classification}
\begin{figure}[h]
\centering
\includegraphics[width=\textwidth]{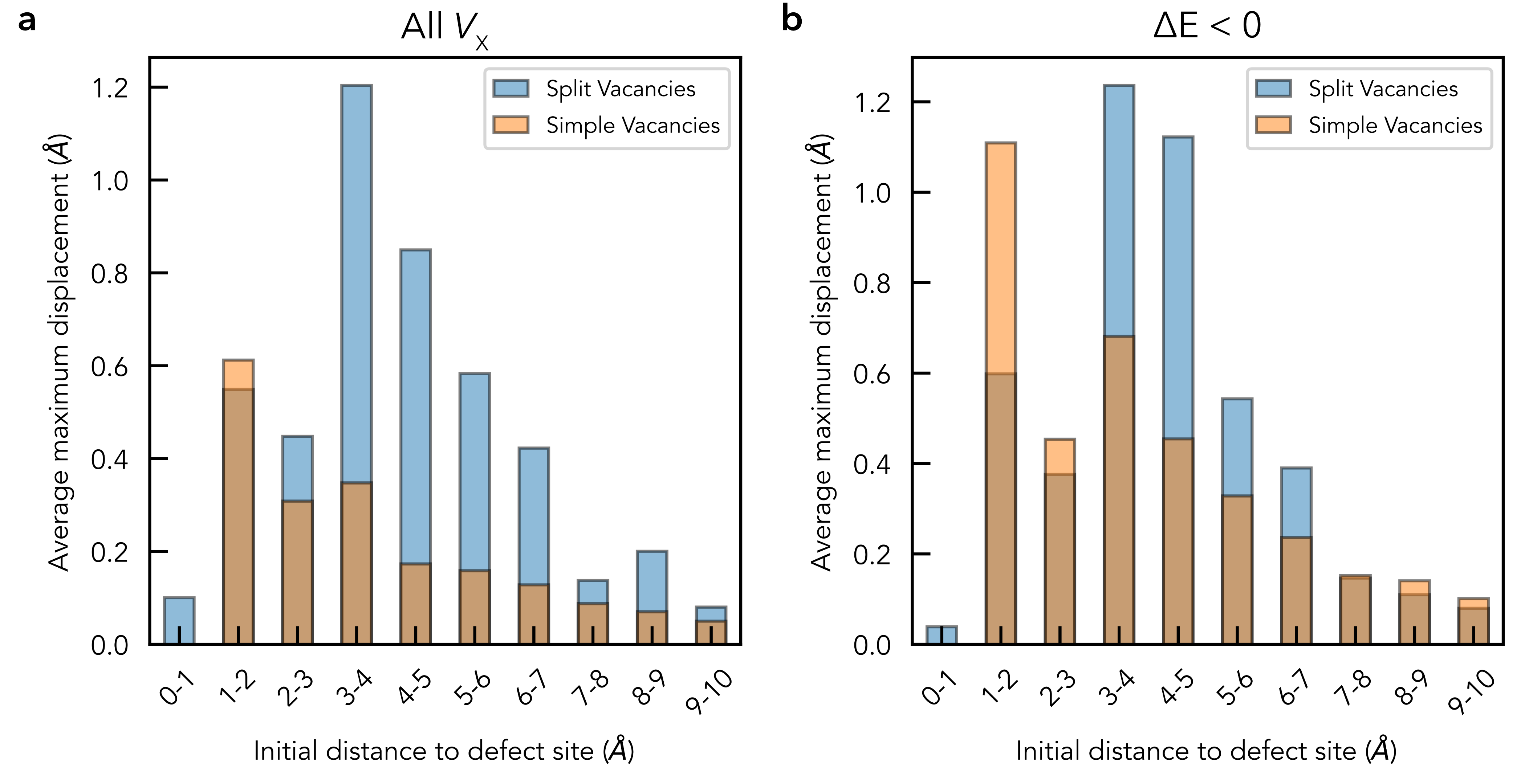}
\caption{Maximum displacement of atoms in relaxed defect structures, relative to their bulk atomic positions, as a function of initial distance to the defect site, averaged over several thousand defect supercell relaxations for cation vacancies in metal oxides.\cite{kumagai_insights_2021} 
Results are plotted separately for relaxations which yielded split vacancy configurations (as classified using the \texttt{doped}\cite{kavanagh_doped_2024} site-matching algorithm; see text), vs simple vacancies.
Average maximum displacements are shown for all cation vacancy relaxations in \textbf{(a)}, while only those for relaxations yielding energy-lowering reconstructions (relative to an unperturbed single vacancy; $\Delta E < 0$) are shown in \textbf{(b)}.}
\label{sifig:displacements}
\end{figure}

In order to classify relaxed defect geometries as split vacancies, simple vacancies or `non-trivial' vacancies,\cite{kumagai_insights_2021} a simple geometric algorithm was employed using the efficient site-matching and structural analysis functions in \texttt{doped},\cite{kavanagh_doped_2024} similar to that employed by \citeauthor{kumagai_insights_2021}\cite{kumagai_insights_2021}
Specifically, split vacancy geometries were characterised as those where 2 sites from the bulk structure cannot be mapped to any site in the defect structure, and 1 site in the defect structure cannot be mapped to any bulk site, within a chosen distance tolerance (as a fraction of the bulk bond length) -- corresponding to the two vacancy and one interstitial sites respectively in the 2\kv{V}{X} + \kv{X}{i} definition of a split vacancy geometry (as discussed in the main text).
A simple vacancy corresponds to cases where 1 site from the bulk structure cannot be matched to the defect structure while all relaxed defect sites can be matched to bulk sites, and `non-trivial' vacancies are all other cases.
In this work, a distance tolerance of \SI{50}{\%} of the bulk bond length was used, however this can be tuned by the user within \texttt{doped}.
The vast majority of relaxed structures in this work are classified as either simple or split vacancies.

\subsection{Oxidation States}
Oxidation states --- used to identify cationic species and compute electrostatic energies --- were determined using the \texttt{doped} algorithms,\cite{kavanagh_doped_2024} which first attempts to use a maximum \textit{a posteriori} estimation approach with bond valences and ICSD oxidation states (as implemented in the \texttt{BVAnalyzer} class in \texttt{pymatgen}),\cite{okeefe_atom_1991,ong_python_2013} then trialling the \texttt{pymatgen}\cite{ong_python_2013} oxidation-state guessing functions (based on ICSD prevalences) if that fails.
Of the $\sim$150,000 compounds in the current Materials Project database, integer oxidation states are determined for $\sim$110,000 compounds.\\

\subsection{Finite-Size Charge Corrections}\label{SI:Finite_Size_Corrections}
Within the supercell approach, the formation energy of a charged defect $X^q$ is defined as:\cite{freysoldt_first-principles_2014}
\begin{align}
\Delta E_\mathrm{f}^{X^q} = E^{X^q} - E^\mathrm{bulk} - \sum_in_i\mu_i +  qE_\mathrm{F} + E_\mathrm{corr} \label{eq:general_formation_energy}
\end{align}
Here, $E^{X^q}$ is the energy of the defect supercell, $E^\mathrm{bulk}$ is the energy of a reference pristine (`bulk') supercell, $\sum_in_i\mu_i$ and $qE_\mathrm{F}$ are atomic and electronic chemical potential terms, and $E_\mathrm{corr}$ is a correction term to account for residual electrostatic interactions arising from the finite-sized supercells.
When comparing split vacancies ($[$\kv{V}{X}-\kv{X}{i}-\kv{V}{X}$]^q$) to point vacancies (\kvc{V}{X}{q}) of the same charge state $q$, the atomic and electronic chemical potential terms $\sum_in_i\mu_i +  qE_\mathrm{F}$ are the same and cancel out.
$E_\mathrm{corr}$, on the other hand, depends on both the charge state (which is the same) and the defect geometry --- or more specifically, the charge distribution in the defect structure, which can differ between the split and point vacancies.
Notably, the magnitude of this finite-size correction --- and thus its potential effect on relative formation energies --- is quadratically dependent on the charge state magnitude ($E_\mathrm{corr} \propto q^2$), inversely dependent on the dielectric constant ($E_\mathrm{corr} \propto \varepsilon^{-1}$), and approximately inverse-polynomially dependent on the supercell length $L$ ($E_\mathrm{corr} \propto aL^{-1} + bL^{-3}$).\cite{freysoldt_fully_2009,kumagai_electrostatics-based_2014}
\\

\begin{figure}[h]
\centering
\includegraphics[width=\textwidth]{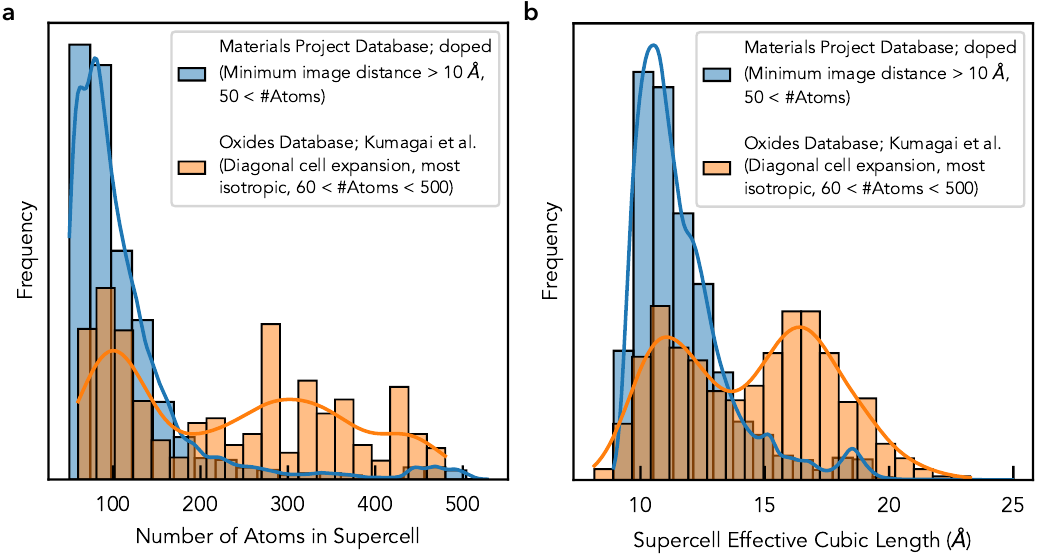}
\caption{Distribution of \textbf{(a)} atom counts and \textbf{(b)} effective cubic lengths for all supercells investigated in this work.
This includes the metal oxides dataset from \citeauthor{kumagai_insights_2021},\cite{kumagai_insights_2021} where supercells were generated from diagonal expansions of conventional unit cells with a minimum/maximum atom count of 60/500 and selecting the most isotropic supercell, and the Materials Project database,\cite{jain_materials_2013} where supercells were generated by searching over all possible primitive cell expansions and taking the smallest supercell with a minimum periodic image distance of \SI{10}
{\angstrom} and a minimum atom count of 50, using the \texttt{doped}\cite{kavanagh_doped_2024} algorithm.
On average, the size and effective cubic lengths of supercells were 230 atoms \& \SI{14.45}{\angstrom}, and 114 atoms \& \SI{11.7}{\angstrom}, for the metal oxides and Materials Project datasets respectively.
}
\label{sifig:Supercell_Size_and_Length_Distributions}
\end{figure}

The effect of the electrostatic finite-size correction on the relative energies of split and point vacancies was analysed for the initial test set of known split vacancy defects shown in \cref{table:split_vacancy_energies}, using the eFNV correction scheme\cite{kumagai_electrostatics-based_2014} from Kumagai \& Oba which natively handles anisotropic dielectric screening and averaging of electrostatic potential shifts across directions.
The centre-of-mass of the split vacancy complexes were used as the defect positions --- which is then used to determine the electrostatic potential sampling region far from the defect site.
Here, the differences in finite-size correction energies between point vacancies and the ground-state split vacancies were found to range between 10 -- \SI{65}{meV} --- corresponding to about \SI{5}{\%} of the finite-size correction magnitude, and $<$\SI{5}{\%} of the point/split vacancy supercell energy differences (\cref{table:split_vacancy_energies}).
However, for some of the metastable split vacancies which have larger \kv{V}{X}-\kv{X}{i} distances and lower symmetry arrangements (thus larger dipole moments and reduced distances between periodic images of \kv{V}{X}/\kv{X}{i}), these differences in the finite-size correction energies for the supercells used were significantly larger, up to $\sim$\SI{10}{\%} of the correction magnitude; \SI{0.1}{eV} for $C2/m$ \ce{Ga2O3} and \SI{0.3}{eV} for $C2/c$ \ce{Sb2O5}.
% \SI{20}{meV} -- \SI{65}{meV} 0.23eV for meta... (or 0.12 eV if tuned...)
% \SI{35}{meV} for Sb2O5, 0.3 eV for meta...
% 5%, 5$, 31%, 2%, 36%, 
In each of these cases, the charge correction was predicted to be lower energy with the eFNV correction scheme.
While these values are still significantly lower than the point/split vacancy supercell energy differences, they do indicate that the impact of finite-size corrections on relative formation energies of point vs split vacancies can be significant, particularly in cases of larger \kv{V}{X}-\kv{X}{i} distances, high charge states, smaller supercells.
As such, evaluating the actual thermodynamic preference for split vs point vacancies in cases of small supercell energy differences requires careful consideration of finite-size corrections (potentially requiring calculations in larger supercells), along with degeneracy effects (discussed in the conclusions), energy functional choices, and potentially other free energy effects.\cite{mosquera-lois_imperfections_2023,arnab_quantitative_2025}
On average, these supercells contained 102 atoms and had effective cubic lengths (i.e. taking the cube root of the supercell volume) of \SI{10.6}{\angstrom}.
The distribution of atom counts and effective cubic lengths for all supercells in the metal oxides and Materials Project datasets investigated in this work are shown in \cref{sifig:Supercell_Size_and_Length_Distributions}.

Formally, finite-size corrections for these defect complexes should account for the presence of multiple charge centres, rather than assuming a single charge centre as for point defects --- however the importance of this treatment decays with larger supercell sizes.
The \texttt{sxdefectalign} (\url{https://sxrepo.mpie.de/projects/sphinx-add-ons/files})\cite{freysoldt_fully_2009} script, which implements the FNV correction and allows the modelling of multiple point charges, was also trialled, however the resulting correction was sensitive to choices of Gaussian model charge width, exponential decay parameter and averaging of alignment constants along different lattice directions (with differences arising from anisotropic defect geometries).

\subsection{Machine Learning Regression}
A machine learning (ML) informatics approach was also trialled for the prediction of split vacancy formation, where ML regression and classification models such as Random Forests, Support Vector Machines, K-Nearest Neighbour and Decision Trees were trained on the dataset of DFT-computed relative energies of split and simple vacancies in metal oxides, using the \texttt{scikit-learn} Python package.\cite{pedregosa_scikit-learn_2011}
Here, sets of simple physical and chemical descriptors were generated for each host structure and candidate split vacancy geometry, such as the distance to the closest lattice site from the vacancy position(s), cation-anion bond lengths, cation valence, orbital subshell ($s$/$p$/$d$) and periodic group, ionic radii, defect charge state etc.
While a number of hyperparameter sweeps, feature selection and regularisation approaches were trialled, these models were found to mostly overfit to the training data.
The code implementing these ML classification models is included in the open-access repository of code and data accompanying this work.

\section{\texttt{ShakeNBreak} applied to \kv{V}{Ga} in $\alpha$-\ce{Ga2O3}}
\begin{figure}[h!]
\centering
\includegraphics[width=\textwidth]{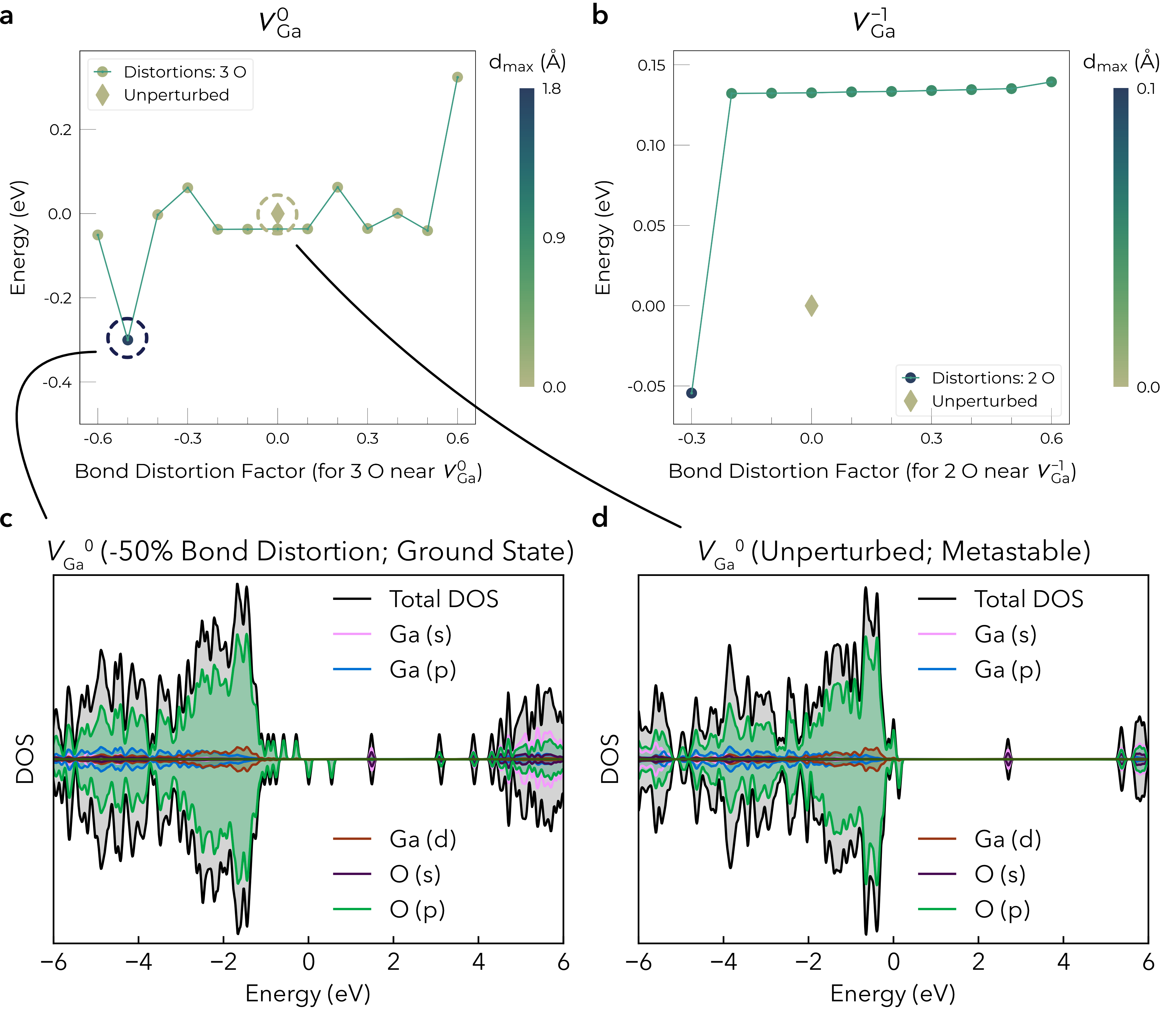}
\caption{\texttt{ShakeNBreak}\cite{mosquera-lois_shakenbreak_2022} structure searching for \kv{V}{Ga} defects in $\alpha$-\ce{Ga2O3} ($R\bar{3}c$). \textbf{a,b} Relative energy versus initial bond distortion factor for trial geometries generated by \texttt{ShakeNBreak}, relaxed using PBEsol semi-local DFT, for \textbf{(a)} \kvc{V}{Ga}{0} and \textbf{(b)} \kvc{V}{Ga}{-1}.
\textbf{c,d} Electronic density of states (DOS) for the \textbf{(c)} ground-state and \textbf{(d)} unperturbed metastable \kvc{V}{Ga}{0} structures.
}
\label{sifig:PBEsol_SnB_R3c_Ga2O3}
\end{figure}

\section{Energy Distributions of Candidate Split Vacancies}
\begin{figure}[h]
\centering
\makebox[\textwidth][c]{
\includegraphics[width=1.25\textwidth]{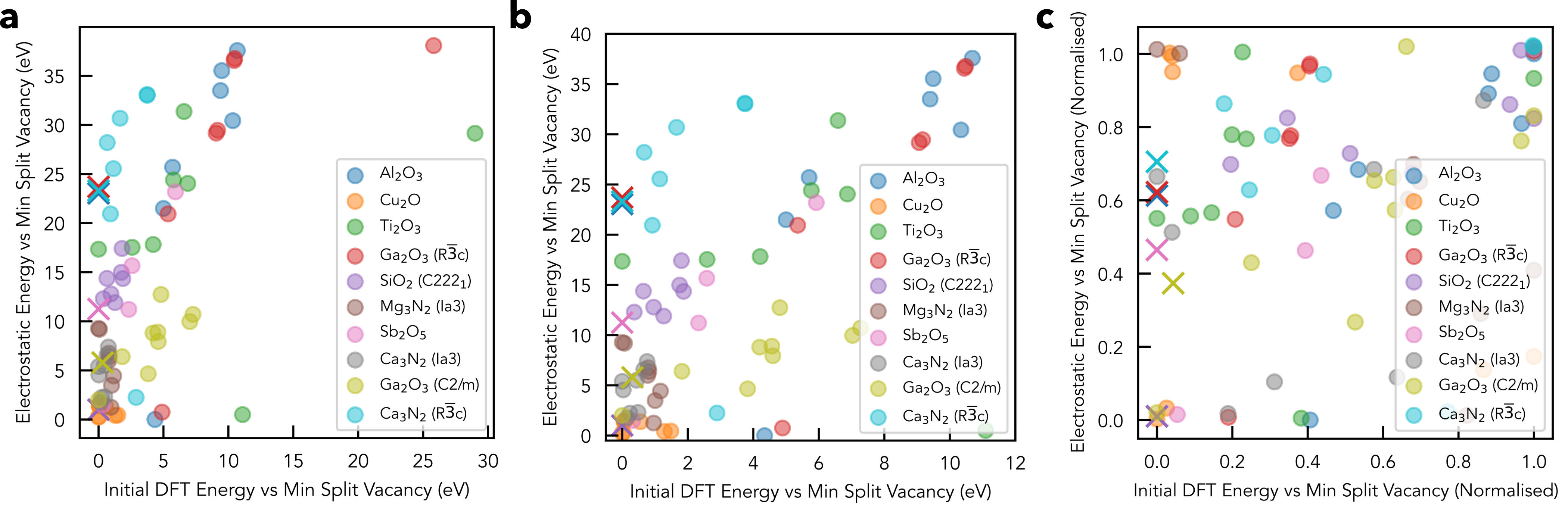}
}
\caption{Electrostatic and DFT energies of initial (un-relaxed) candidate split vacancy configurations relative to the minimum energy (un-relaxed) split vacancy structure, across the same initial compound test set as \cref{fig:Geometric_and_PBE0}b (\kv{V}{Cation} for oxides and \kv{V}{Anion} for nitrides).
\textbf{(a)} and \textbf{(b)} show the energy distributions over different DFT energy ($x$-axis) ranges, while \textbf{(c)} shows the energy distributions normalised by the energy range for each compound.
}
\label{sifig:Initial_Test_Set_SI_ES_Analysis}
\end{figure}

\begin{figure}[h]
\centering
\includegraphics[width=0.7\textwidth]{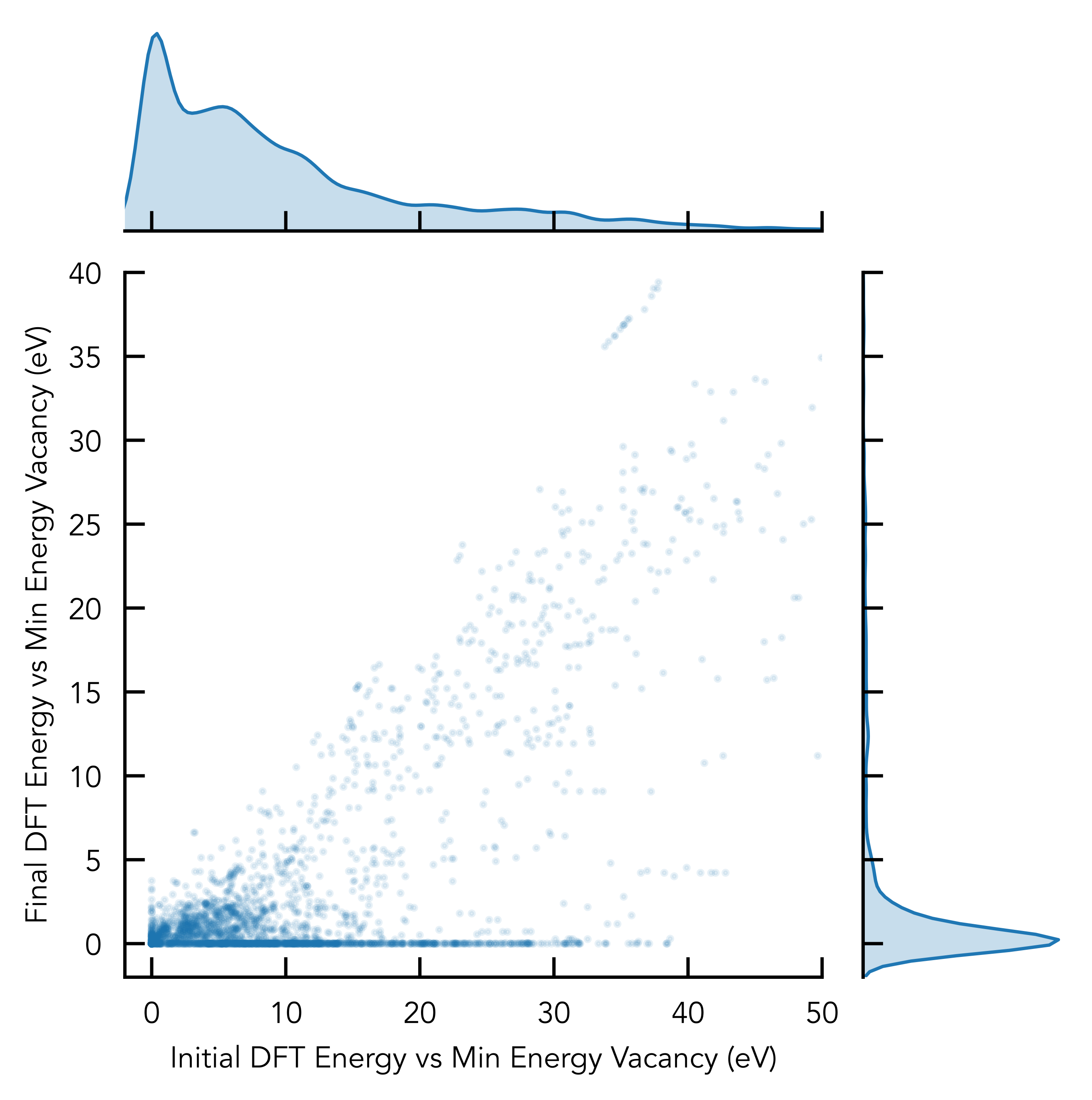}
\caption{Joint distribution plot of the final vs initial DFT energies of all candidate split vacancies in the full DFT calculated dataset ($\sim$1000 compounds), relative to the minimum energy candidate split vacancy geometry.
`Initial' refers to the fact that these energies are computed for candidate split vacancies before performing geometry relaxation (as in the electrostatic screening step).
}
\label{sifig:Initial_vs_Final_DFT_energies}
\end{figure}

\begin{figure}[h]
\centering
\includegraphics[width=0.7\textwidth]{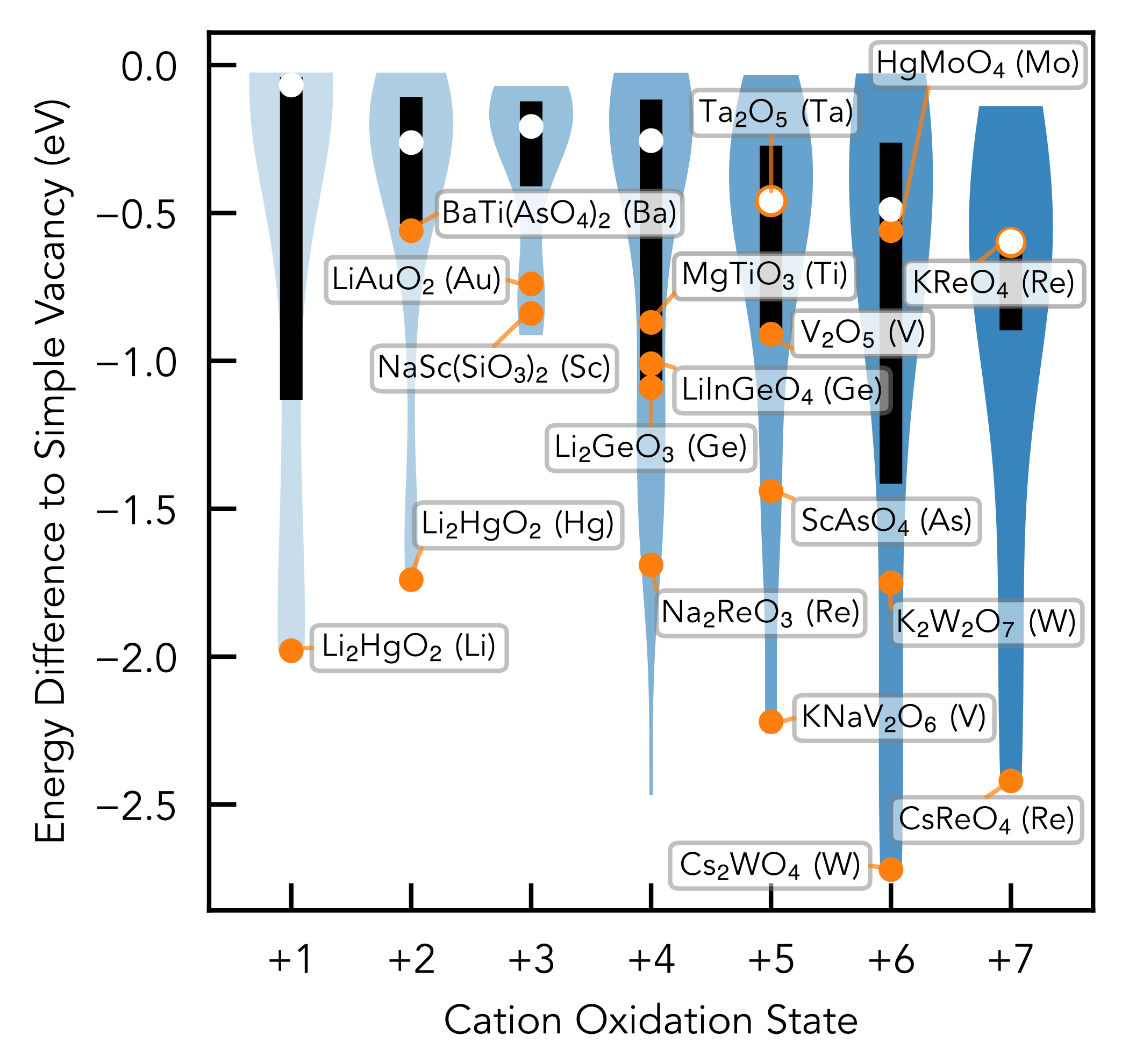}
\caption{Distribution of energies for all lower energy structures (regardless of split vacancy classification) relative to the lowest energy symmetry-inequivalent point vacancy for different cation oxidation states, for all lower energy vacancies in metal oxides identified in this work ($\Delta E < \SI{-0.025}{eV}$). Some example compounds and the corresponding cation (vacancy) are shown as labelled orange datapoints. White circles and black rectangles denote the median and inter-quartile range respectively.
}
\label{sifig:Kumagai_Oxides_Energy_Lowering_Labelled}
\end{figure}

\clearpage

\section{\texttt{MACE-mp} Foundation Model Geometry Optimisation Tests with ASE\cite{larsen_atomic_2017}}\label{SI:MACE_tests}
\begin{figure}[h]
\centering
\includegraphics[width=\textwidth]{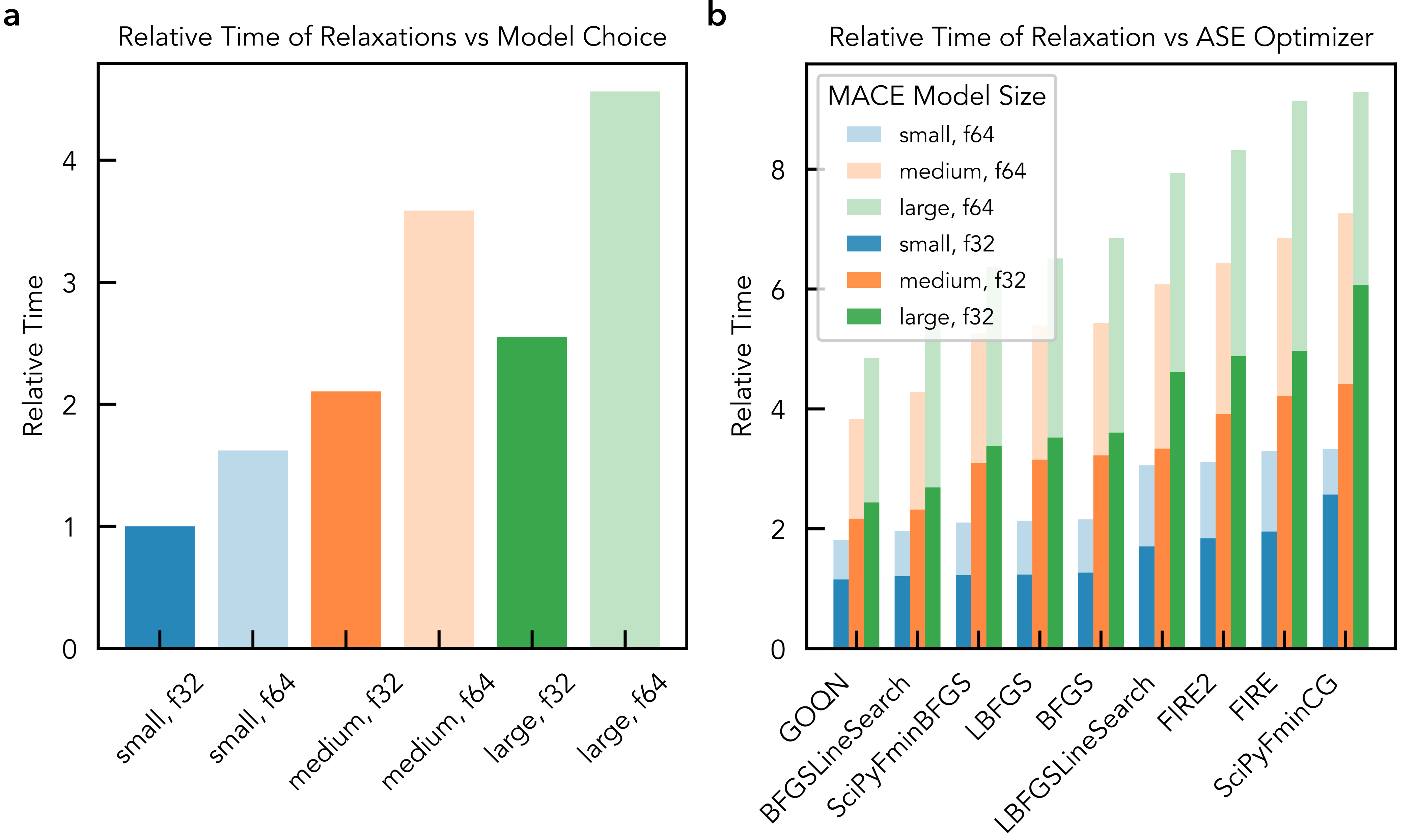}
\caption{Mean relative runtimes of all \texttt{MACE-mp} geometry optimisations as a function of model size and float precision, averaged over all force minimisation algorithms \textbf{(a)}, and separated by optimisation algorithm \textbf{(b)}.
}
\label{sifig:MACE_Float_Size_Tests}
\end{figure}

For \texttt{MACE-mp} model size, \texttt{small} and \texttt{large} were found to perform similarly in terms of predictive accuracy, with \texttt{medium} being about \SI{20}{\%} worse.
Given the lower runtimes of \texttt{small} models (\cref{sifig:MACE_Float_Size_Tests}) and similar accuracy, the \texttt{small} model was then used for ML-accelerated screening.
The recently-released \texttt{MACE-mp-0b2} model was also tested, giving essentially the same results with a $\sim$\SI{10}{\%} speed increase.
32 and 64-bit precision were both tested and showed similar energy accuracies, with a mean absolute deviation of total energies of \SI{1.4}{meV}, and a standard deviation of absolute differences of \SI{1.1}{meV}.
32-bit precision was found to be $\sim$\SI{75}{\%} faster on average and so was used for production-run \texttt{MACE-mp} relaxations.\\
% Check through notebooks and add info about relevant points / tested things (e.g. supercell size dependence, ML models..., VIV distances squared)

\begin{figure}[h]
\centering
\includegraphics[width=\textwidth]{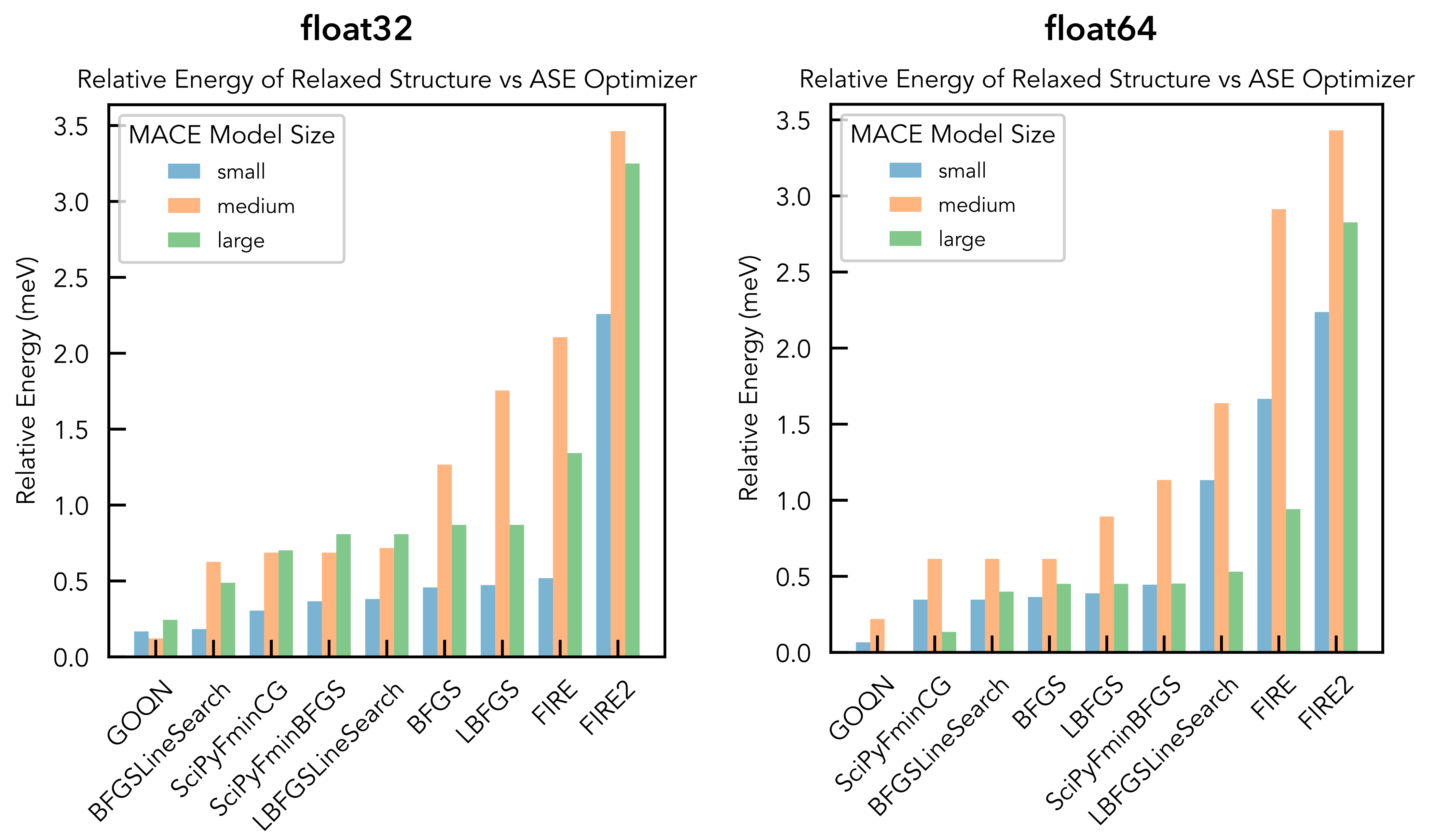}
\caption{Mean final energies of all \texttt{MACE-mp} geometry optimisations, relative to the lowest energy found for the same input geometry, model size and precision (i.e. out of all optimisation algorithms), as a function of force minimisation algorithm and model size.
Results for 32 and 64-bit precisions shown on the left and right respectively.
}
\label{sifig:MACE_Tests_Rel_Energies}
\end{figure}

\begin{figure}[h]
\centering
\includegraphics[width=\textwidth]{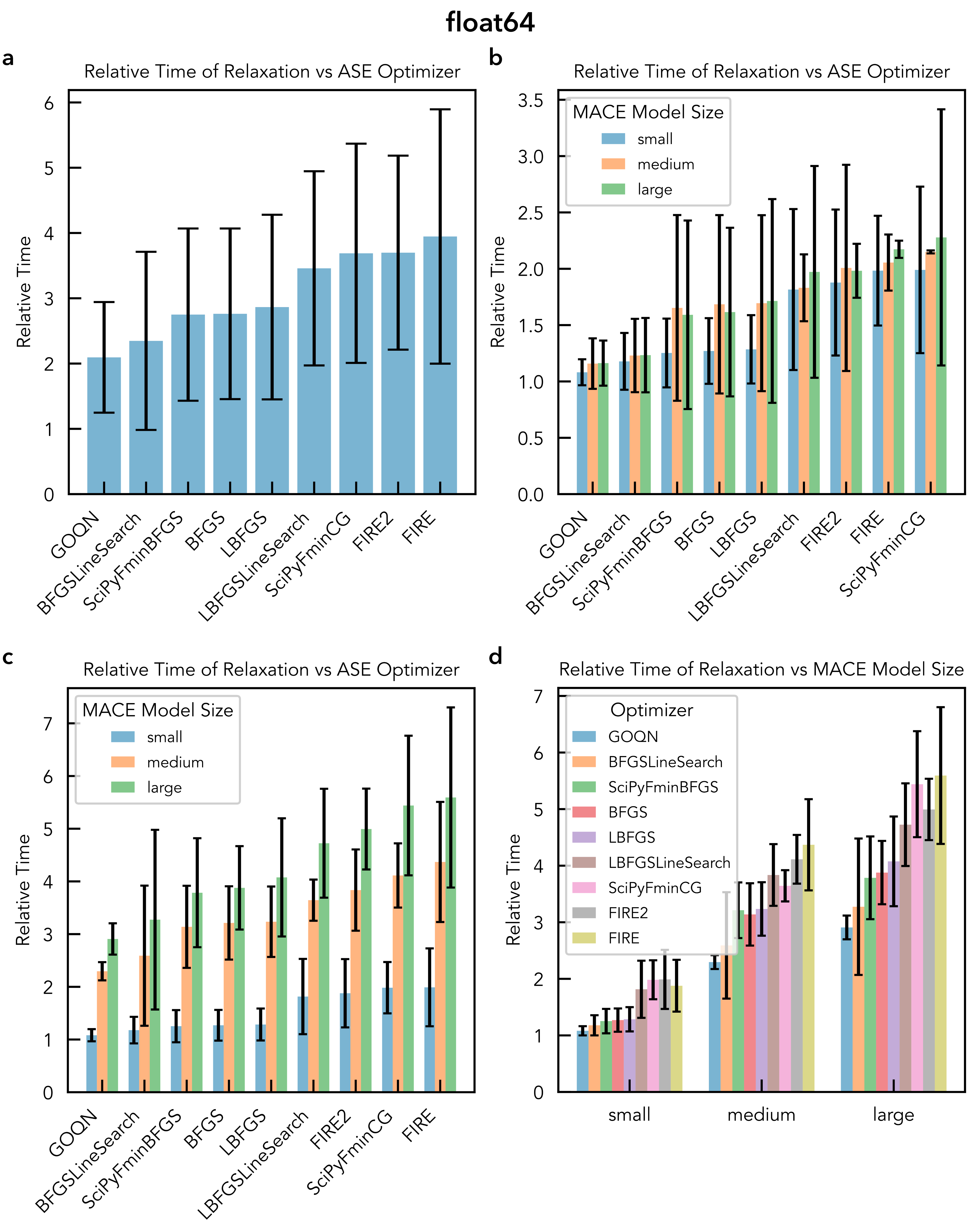}
\caption{Mean relative runtimes of all \texttt{MACE-mp} geometry optimisations with 64-bit precision, as a function of optimisation algorithm \textbf{(a)}, optimisation algorithm and model size (normalised within each model size)\textbf{(b)}, optimisation algorithm and model size (normalised to the \texttt{small} runtimes)\textbf{(c)}, and grouped by model size (again normalised to the \texttt{small} runtimes)\textbf{(d)}.
}
\label{sifig:MACE_f64_Speed_Tests}
\end{figure}

\begin{figure}[h]
\centering
\includegraphics[width=\textwidth]{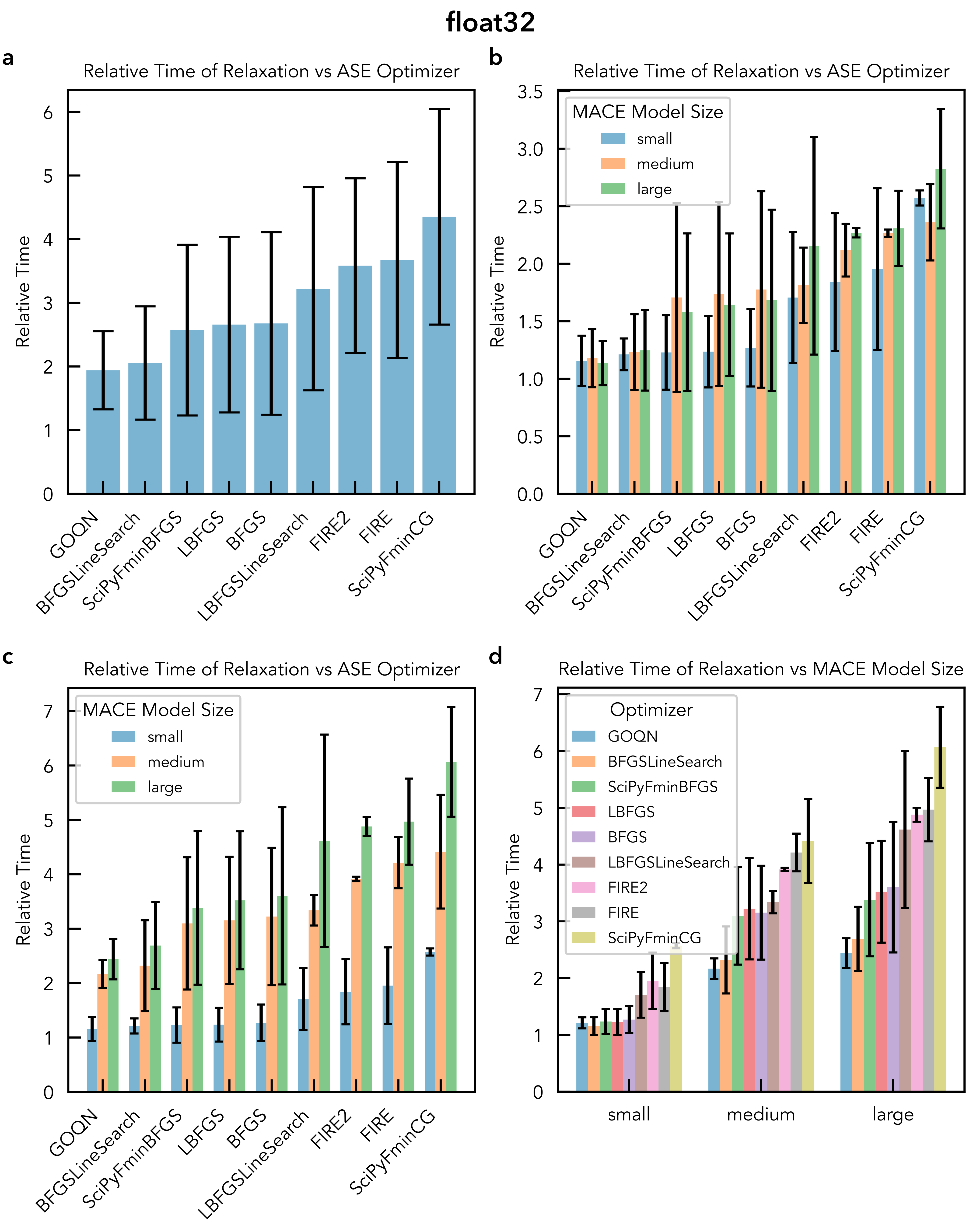}
\caption{Mean relative runtimes of all \texttt{MACE-mp} geometry optimisations with 32-bit precision, as a function of optimisation algorithm \textbf{(a)}, optimisation algorithm and model size (normalised within each model size)\textbf{(b)}, optimisation algorithm and model size (normalised to the \texttt{small} runtimes)\textbf{(c)}, and grouped by model size (again normalised to the \texttt{small} runtimes)\textbf{(d)}.
}
\label{sifig:MACE_f32_Speed_Tests}
\end{figure}

For the \texttt{MACE-mp} geometry optimisation tests shown in \cref{sifig:MACE_f32_Speed_Tests,sifig:MACE_f64_Speed_Tests,sifig:MACE_Float_Size_Tests,sifig:MACE_Tests_Rel_Energies}, relaxations were performed for all candidate split vacancy geometries in the metal oxides test set (\cref{fig:Kumagai_Oxides_Screening}) which were calculated with DFT, corresponding to $\sim$4,000 supercell relaxations.
The \texttt{GOQN} (`good old quasi Newton') optimisation algorithm, implemented in the \texttt{ASE}\cite{larsen_atomic_2017} package was found to be the most accurate and fastest force minimisation algorithm, and so was used for \texttt{MACE-mp} geometry optimisations in the Materials Project\cite{jain_materials_2013} screening.
The Gaussian Process Minimiser (\texttt{GPMin}) algorithm was also trialled, however the high memory demand due to the large supercell sizes caused relaxations to crash.

\begin{figure}[h]
\centering
\includegraphics[width=\textwidth]{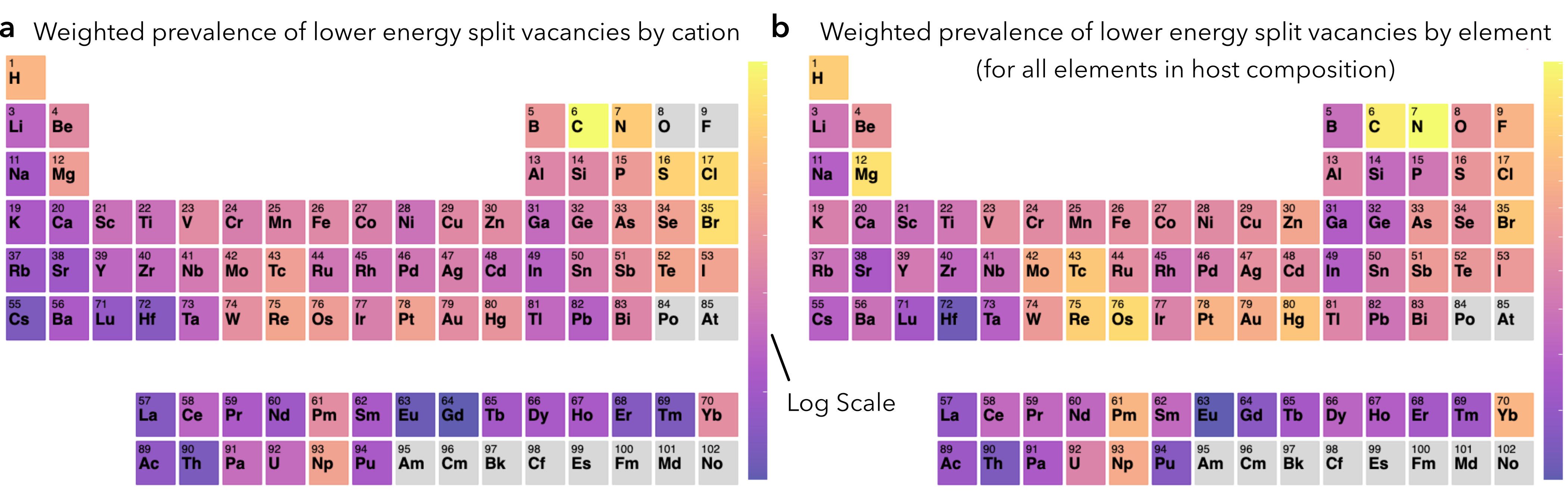}
\caption{Heatmap plots of the normalised prevalence of lower energy split vacancy states predicted by the ML-accelerated screening of the MP database, for \textbf{(a)} the cation vacancy element and \textbf{(b)} all elements in the corresponding host compound, now weighted by the magnitude of the energy lowering and using a logarithmic scale for the colourbar.
Values are normalised by the total prevalence of each element within the dataset.
}
\label{sifig:Weighted_Prevalence_Plots}
\end{figure}

\clearpage

\section{Identified Metastable States}\label{SI:Meta_States}
As mentioned in the main text, the electrostatic screening for low-energy split vacancies (without machine-learned potentials) in the metal oxides dataset\cite{kumagai_insights_2021} identifies many low-energy metastable states, finding 210 distinct metastable states with energies within \SI{0.5}{eV} of the lowest energy simple point vacancy, in 160 of the 600 cation vacancies which gave candidate low-energy sites from electrostatic screening in this test set.
Here. distinct metastable states are those which (i) relax to a split vacancy geometry (determined by the \texttt{doped}\cite{kavanagh_doped_2024} classification algorithm), with no corresponding symmetry-inequivalent point vacancy spontaneously relaxing to a split vacancy, and (ii) are different in energy by $>$\SI{25}{meV} to all other metastable states for that vacancy.
If we loosen criterion (i) to include relaxed geometries which are not classified as split vacancies but have a difference in energy $>$\SI{50}{meV} (but $<$\SI{500}{meV}) to any corresponding symmetry-inequivalent point vacancy --- well above the energy noise in these calculations --- this increases to 300 distinct metastable states in 200 of the 600 cation vacancies with candidate low-energy sites from electrostatic screening.

\begin{figure}[h]
\centering
\includegraphics[width=\textwidth]{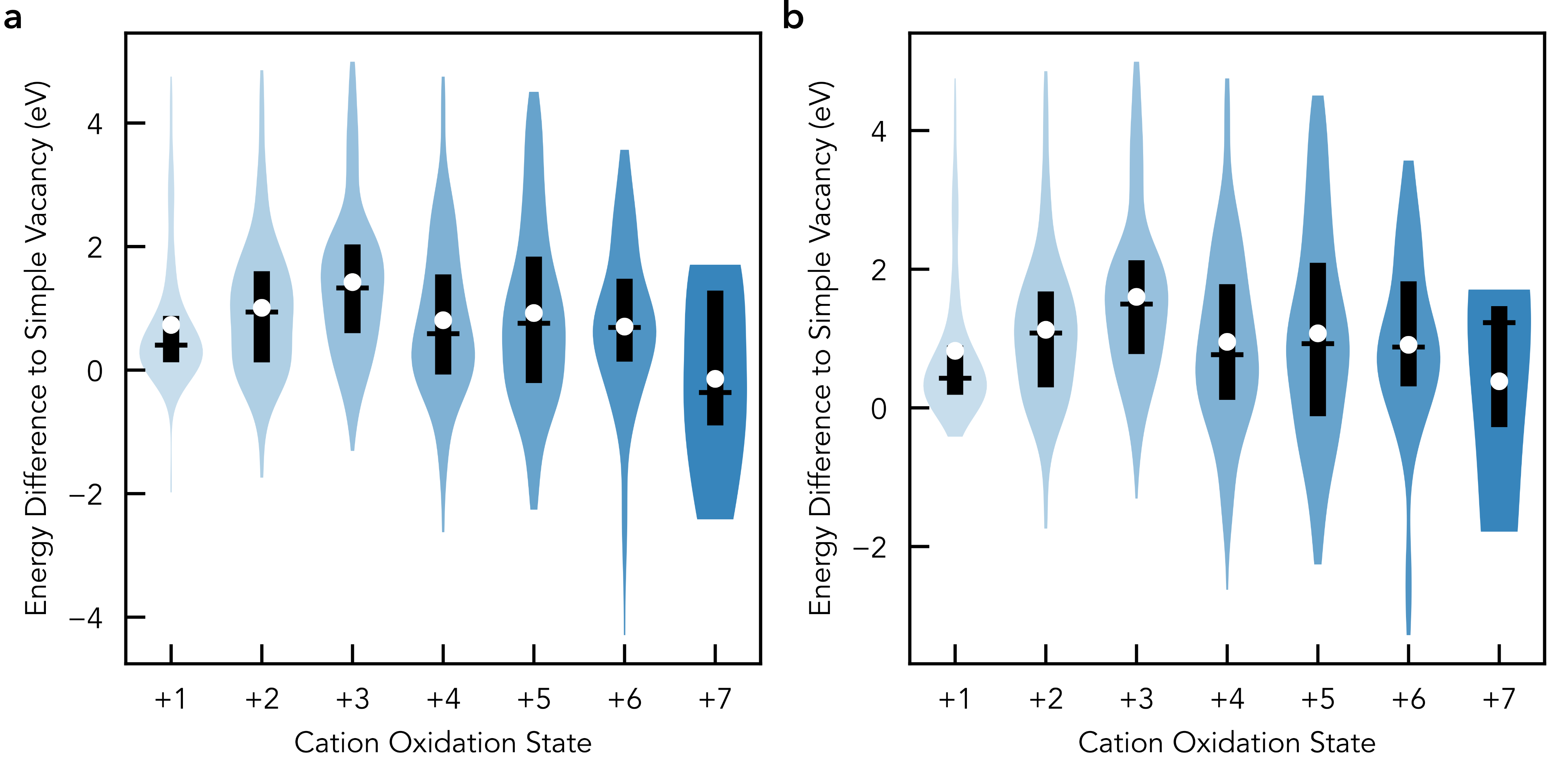}
\caption{
\textbf{(a)} Energy distribution of all distinct metastable states relative to the lowest energy symmetry-inequivalent point vacancy for different cation oxidation states, for all metal oxides calculated in this work. 
Distinct metastable states are classified as those which (i) relax to a split vacancy geometry (determined by the \texttt{doped}\cite{kavanagh_doped_2024} classification algorithm), with no corresponding symmetry-inequivalent point vacancy spontaneously relaxing to a split vacancy, or have a difference in energy $>$\SI{50}{meV} to any corresponding symmetry-inequivalent point vacancy, and (ii) are different in energy by $>$\SI{25}{meV} to all other metastable states for that vacancy.
Subfigure \textbf{(b)} shows the distributions when \emph{only} split vacancy geometries are included.
White circles, black dashes and rectangles denote the mean, median and inter-quartile range respectively.
}
\label{sifig:All_Oxides_Distinct_Meta_States_vs_Cation_Oxi}
\end{figure}

\begin{scriptsize}
\begin{longtable}{l|c|c|c|c}
% \label{sitab:All_Distinct_Meta_Oxides_0pt5}
\caption{Energies of all distinct metastable states relative to the lowest energy symmetry-inequivalent point vacancy ($\Delta E$), with $\Delta E < \SI{0.5}{eV}$, for all metal oxides calculated using DFT in this work. 
Distinct metastable states are classified as those which (i) relax to a split vacancy geometry, with no corresponding symmetry-inequivalent point vacancy spontaneously relaxing to a split vacancy, or have a difference in energy $>$\SI{50}{meV} to any corresponding symmetry-inequivalent point vacancy, and (ii) are different in energy by $>$\SI{25}{meV} to all other metastable states for that vacancy.\label{sitab:All_Distinct_Meta_Oxides}} \\
\textbf{Formula}                    & \textbf{Cation} & 
\begin{tabular}[t]{@{}c@{}}\textbf{Space} \\ \textbf{Group} \end{tabular} & 
\begin{tabular}[t]{@{}c@{}}\textbf{Energy} \\ \textbf{(eV)} \end{tabular} & 
\begin{tabular}[t]{@{}c@{}}\textbf{Split} \\ \textbf{Vacancy} \end{tabular} \\
\hline
\endhead
Ag$_{2}$BiO$_{3}$                   & Ag     & Pnn2                           & 0.14        & True          \\
Al$_{2}$O$_{3}$                     & Al     & C2/c                           & -0.58       & True          \\
Al$_{2}$O$_{3}$                     & Al     & C2/m                           & -1.31       & True          \\
Al$_{2}$O$_{3}$                     & Al     & C2/m                           & -0.68       & True          \\
Al$_{2}$O$_{3}$                     & Al     & C2/m                           & 0.33        & True          \\
Al$_{2}$O$_{3}$                     & Al     & R$\overline{3}$c               & -0.58       & True          \\
AlSbO$_{4}$                         & Sb     & Cmmm                           & 0.45        & True          \\
AlTl(MoO$_{4}$)$_{2}$               & Tl     & P$\overline{3}$m1              & -0.06       & False         \\
\end{longtable}
\end{scriptsize}

\begin{scriptsize}
\begin{longtable}{l|c|c|c|c}
\caption{Energies of all distinct metastable states relative to the lowest energy symmetry-inequivalent point vacancy ($\Delta E$), with $\Delta E < \SI{0.5}{eV}$, for all nitrides calculated using DFT in this work. 
Distinct metastable states are classified as those which (i) relax to a split vacancy geometry, with no corresponding symmetry-inequivalent point vacancy spontaneously relaxing to a split vacancy, or have a difference in energy $>$\SI{50}{meV} to any corresponding symmetry-inequivalent point vacancy, and (ii) are different in energy by $>$\SI{25}{meV} to all other metastable states for that vacancy.\label{sitab:All_Distinct_Meta_Nitrides}} \\
\textbf{Formula}                    & \textbf{Cation} & 
\begin{tabular}[t]{@{}c@{}}\textbf{Space} \\ \textbf{Group} \end{tabular} & 
\begin{tabular}[t]{@{}c@{}}\textbf{Energy} \\ \textbf{(eV)} \end{tabular} & 
\begin{tabular}[t]{@{}c@{}}\textbf{Split} \\ \textbf{Vacancy} \end{tabular} \\
\hline
\endhead

Ag$_{6}$HgNO$_{11}$                       & Ag     & Fmm2                           & -0.32       & False         \\
Ag$_{7}$NO$_{11}$                         & Ag     & Fmm2                           & 0.12        & False         \\
AlC$_{3}$(NCl$_{2}$)$_{3}$                & Al     & P$\overline{1}$                & 0.06        & True          \\
Ba$_{3}$BiN                               & Ba     & P6$_{3}$/mmc                   & -0.08       & False         \\
Ba$_{3}$Si$_{6}$(NO$_{6}$)$_{2}$          & Si     & P$\overline{3}$                & -0.47       & False         \\
Ba$_{5}$Re$_{3}$NO$_{18}$                 & Ba     & P6$_{3}$cm                     & 0.09        & False         \\
Ba(Si$_{3}$N$_{4}$)$_{2}$                 & Si     & Imm2                           & -0.67       & False         \\
Ca$_{2}$MoN$_{3}$                         & Mo     & C2/c                           & -0.59       & False         \\
\end{longtable}
\end{scriptsize}

% \putbib  % SI bibliography
% \end{bibunit}
\end{document}